\documentclass[a4paper, 12pt]{article}
\usepackage{jheppub}
\usepackage[utf8x]{inputenc}
\usepackage{slashed}
\usepackage{xcolor}
\usepackage{caption}
\usepackage{subcaption}
\usepackage[T1]{fontenc}
\usepackage[english]{babel}
\usepackage{amsmath}
\usepackage[english]{babel}
\usepackage{nicefrac}

\usepackage{graphicx}
\usepackage{float}
\usepackage{hyperref}

\usepackage{graphicx}
\usepackage{tikz}
\usepackage[com pat=1.1.0]{tikz-feynman}
\usepackage{circuitikz}
\usetikzlibrary{decorations.pathmorphing}
\usetikzlibrary{shapes.geometric}
\usetikzlibrary{matrix, calc, arrows}
\tikzset{snake it/.style={decorate, decoration=snake}}
\tikzset{gluon/.style={decorate,
 decoration={coil,amplitude=3pt, segment length=4pt,  pre length=.1cm, post length=.01cm}}
 }

\usetikzlibrary{decorations.markings}

\tikzset{%
  dots/.style args={#1per #2}{%
    line cap=round,
    dash pattern=on 0 off #2/#1
  }
}

\tikzfeynmanset{
    sdot/.style={
        /tikzfeynman/dot,
        /tikz/minimum size=3pt,
    },
    gluon/.style={decorate,
        decoration={coil,amplitude=3pt, segment length=4pt,  pre length=.1cm, post length=.01cm}
    },
    graviton/.style={decorate, double,
        decoration={snake}
    },
    compact arrow/.style={
        /tikzfeynman/momentum/.cd, arrow distance=8pt, label distance=-6pt, arrow shorten=0.27
    },
}

\newcommand{\be }{\begin{equation}}
\newcommand{\ee }{\end{equation}}
\newcommand{\ba }{\begin{equation} \begin{aligned}}
\newcommand{\ea }{ \end{aligned} \end{equation} }
\newcommand{\la}{\langle}
\newcommand{\ra}{\rangle}

\newcommand{\D}{{\cal D}}
\newcommand{\deltahat}{\hat{\delta}}
\newcommand{\s }{ {\cal S} }
\newcommand{\ccdot}{\! \cdot \!}
\newcommand{\half}{{\textstyle\frac{1}{2}}}
\newcommand{\dd}{\mathrm{d}}

\newcommand{ \wprop} {
	\begin{tikzpicture}[thick]
	\coordinate (A) at (-1,-0);
	\coordinate (B) at (1,-0);
	\filldraw (A) circle (2pt) node[left] {$z^\mu$};
	\filldraw (B) circle (2pt) node[right] {$z^\nu$};
	\draw[] (0,.5) node[]{$\omega$};
	\path[ very thick,draw] (-1,0)--(1,0); 	\path[ thick,draw,->] (-0.5,0.3)--(0.5,0.3) node[pos=.5] {}; 
	\end{tikzpicture}
}

\newcommand{\sprop} {
    \begin{tikzpicture}[thick]
        \path [very thick, draw]
        (-1,-1) -- (1,-1) node[currarrow,scale=1,xscale=1,sloped,pos=.5]{};
        \draw[] (-0,-.7) node[] {$\omega$};

        \coordinate (A) at (-1,-1);
        \coordinate (B) at (1,-1);
        \filldraw (A) circle (1.5pt)node[left]{$\Psi_{\mu}$};
        \filldraw (B) circle (1.5pt)node[right]{$\Psi_{\nu}$};
    \end{tikzpicture}
}

\newcommand{ \cprop} {
    \begin{tikzpicture}[thick]
        \path [very thick, draw]
        (-1,-1) -- (1,-1) node[currarrow,scale=1,xscale=1,sloped,pos=.5]{};
        \draw[] (-0,-.7) node[] {$\omega$};

        \coordinate (A) at (-1,-1);
        \coordinate (B) at (1,-1);
        \filldraw (A) circle (1.5pt)node[left]{$\lambda_{\alpha}$};
        \filldraw (B) circle (1.5pt)node[right]{$\bar{\lambda}_{\beta}$};
    \end{tikzpicture}
}

\newcommand{ \gluprop} {
    \begin{tikzpicture}[thick]
        \draw [gluon]
        (-1,-1) -- (1,-1);
        \draw[->,>=stealth] (-.6,-.7) -- (.6,-.7) node[above,midway]
        {$q $};
        \coordinate (A) at (-1,-1);
        \coordinate (B) at (1,-1);
        \filldraw (A) circle (1.5pt)node[left]{$A^{a}_{\mu}$};
        \filldraw (B) circle (1.5pt)node[right]{$A^{b}_{\nu}$};
    \end{tikzpicture}
}

\newcommand{ \grprop} {
    \begin{tikzpicture}[thick]
        \draw [double,snake it]
        (-1,-1) -- (1,-1);
        \draw[->,>=stealth] (-.6,-.7) -- (.6,-.7) node[above,midway]
        {$q $};
        \coordinate (A) at (-1,-1);
        \coordinate (B) at (1,-1);
        \filldraw (A) circle (1.5pt)node[left]{$h_{\mu\nu}$};
        \filldraw (B) circle (1.5pt)node[right]{$h_{\rho\sigma}$};
    \end{tikzpicture}
}

\newcommand{ \vglu} {\begin{tikzpicture}[thick]
        \path [dashed, draw]
        (-1,-1) -- (1,-1)node[right]
        {$\quad $};
        \path [ draw,gluon]
        (0,-1) -- (0,-2)node[below]{$A_{\mu}^{a}(q)$};
        \draw[->,>=stealth] (-0.5,-1.2) -- (-0.5,-1.8) node[midway, left]
        {$q$};
        \coordinate (A) at (0,-1);
        \filldraw (A) circle (1.5pt);
    \end{tikzpicture}
}

\newcommand{ \vtwoglu} {\begin{tikzpicture}[thick]
        \path [dashed, draw]
        (-1,-1) -- (1,-1)node[right]
        {$\quad $};
        \path [ draw,gluon]
        (0,-1) -- (-1,-2)node[below]{$A_{b\mu}(q_{1})$};
        \path [ draw,gluon]
        (0,-1) -- (1,-2)node[below]{$A_{c\nu}(q_{2})$};
        \coordinate (A) at (0,-1);
        \filldraw (A) circle (1.5pt);
    \end{tikzpicture}
}

\newcommand{\vKinglu}{\begin{tikzpicture}[thick]
        \path [dashed, draw]
        (-1,-1) -- (0,-1);
        \path [very thick, draw]
        (0,-1) -- (1,-1)node[right]
        {$z_\alpha(\omega) $};
        \path [ draw,gluon]
        (0,-1) -- (0,-2)node[below]{$A_{\mu}^{a}(q)$};
        \draw[->,>=stealth] (-0.5,-1.2) -- (-0.5,-1.8) node[midway, left]
        {$q$};
         \draw[->,>=stealth] (0.3,-.7) -- (1,-.7) node[above,midway]
        {$\omega$};

        \coordinate (A) at (0,-1);
        \filldraw (A) circle (1.5pt);
    \end{tikzpicture}
}

\newcommand{\vSpinglu }{\begin{tikzpicture}[thick]
        \path [dashed, draw]
        (-1,-1) -- (0,-1);
        \path [very thick, draw]
        (0,-1) -- (1,-1)node[right]
        {$\Psi^\nu(\omega) $};
        \path [very thick, draw]
        (0,-1) -- (1,-1)node[currarrow,scale=1,xscale=1,sloped,pos=.5]{};
        \path [ draw,gluon]
        (0,-1) -- (0,-2)node[below]{$A_{\mu}^{a}(q)$};
        \draw[->,>=stealth] (-0.5,-1.2) -- (-0.5,-1.8) node[midway, left]
        {$q$};
        \coordinate (A) at (0,-1);
        \filldraw (A) circle (1.5pt);
    \end{tikzpicture}
}

\newcommand{\vColglu}{\begin{tikzpicture}[thick]
        \path [dashed, draw]
        (-1,-1) -- (0,-1);
        \path [very thick, draw]
        (0,-1) -- (1,-1)node[right]
        {$\lambda^\alpha(\omega) $};
        \path [very thick, draw]
        (0,-1) -- (1,-1)node[currarrow,scale=1,xscale=1,sloped,pos=.5]{};
        \path [ draw,gluon]
        (0,-1) -- (0,-2)node[below]{$A_{\mu}^{a}(q)$};
        \draw[->,>=stealth] (-0.5,-1.2) -- (-0.5,-1.8) node[midway, left]
        {$q$};
        \coordinate (A) at (0,-1);
        \filldraw (A) circle (1.5pt);
    \end{tikzpicture}
}

\newcommand{\vbarColglu}{\begin{tikzpicture}[thick]
    \path [dashed, draw]
    (-1,-1) -- (0,-1);
    \path [very thick, draw]
    (0,-1) -- (1,-1)node[right]
    {$\bar{\lambda}_\alpha(\omega) $};
    \path [very thick, draw]
    (0,-1) -- (1,-1)node[currarrow,scale=1,xscale=-1,sloped,pos=.5]{};
    \path [ draw,gluon]
    (0,-1) -- (0,-2)node[below]{$A_{\mu}^{a}(q)$};
    \draw[->,>=stealth] (-0.5,-1.2) -- (-0.5,-1.8) node[midway, left]
    {$q$};
    \coordinate (A) at (0,-1);
    \filldraw (A) circle (1.5pt);
\end{tikzpicture}
}

\def \vaxion {\begin{tikzpicture}[thick]
        \path [dashed, draw]
        (-1,-1) -- (1,-1)node[right]
        {$\quad $};
        \path [ draw,snake it]
        (0,-1) -- (0,-2)node[below]{$B_{\mu\nu}$};
        \draw[->,>=stealth] (-0.5,-1.2) -- (-0.5,-1.8) node[midway, left]
        {$q$};
        \coordinate (A) at (0,-1);
        \filldraw (A) circle (1.5pt);
    \end{tikzpicture}
}

\newcommand{ \vtwoGR} {\begin{tikzpicture}[thick]
        \path [dashed, draw]
        (-1,-1) -- (1,-1)node[right]
        {$\quad $};
        \path [ draw,double,snake it]
        (0,-1) -- (-1,-2)node[below]{$h_{\mu\nu}(q_{1})$};
        \path [ draw,double,snake it]
        (0,-1) -- (1,-2)node[below]{$h_{\rho\sigma}(q_{2})$};
        \coordinate (A) at (0,-1);
        \filldraw (A) circle (1.5pt);
    \end{tikzpicture}
}

\def \vhhh { \begin{tikzpicture}[thick]
        \path[draw,double,snake it] (0,0)--(0,.8)node[above]{$h_{\mu\nu}(q_1)$};
        \path[draw,double,snake it] (0,0)--(-0.7,-0.7)node[left]{$h_{\rho\sigma}(q_2)$};
        \path[draw,double,snake it] (0,0)--(0.7,-0.7)node[right]{$h_{\alpha\beta}(q_3)$};

        \coordinate (A) at (0,0);
        \filldraw (A) circle (1.5pt);
    \end{tikzpicture}
}

\newcommand{\loeikYM} { \begin{tikzpicture}[thick]
        \path [dashed,draw]
        (-1,-.5) node[left]{1} --(1,-.5);
        \path [dashed,draw]
        (-1,-2) node[left]{2} --(1,-2);
        \path [draw,gluon]
        (0,-.5) -- (0,-2);
        \draw[<-,>=stealth] (-0.5,-1) -- (-0.5,-1.6) node[midway, left]
        {$q_2$};
        \coordinate (A) at (0,-.5);
        \filldraw (A) circle (1.5pt);
        \coordinate (B) at (0,-2);
        \filldraw (B) circle (1.5pt);
    \end{tikzpicture}
}

\newcommand{\loradz}[3] {
    \begin{tikzpicture}[thick]
        \path [dashed,draw]
        (-1,-.5) node[left]{1} --(0,-.5) node[above]{$#1 $};

        \path [very thick,draw,#2]
        (0,-.5)--(1,-.5) node[above]{$#3 $};

        \path [dashed,draw]
        (1,-.5)--(2,-.5) node[right]{$\quad $};

        \path [dashed,draw]
        (-1,-2) node[left]{2} --(2,-2);

        \path [ draw,gluon]
        (0,-.5) -- (0,-2);

        \path [ draw,gluon]
        (1,-.5) -- (2,-1.5) node[below,right]{$k$};

        \draw[<-,>=stealth] (-0.5,-1) -- (-0.5,-1.6) node[midway, left]
        {$q_{2}$};

        \draw[->,>=stealth] (.8,-.8) -- (1.5,-1.5);

        \coordinate (A) at (0,-.5);
        \filldraw (A) circle (1.5pt);
        \coordinate (B) at (0,-2);
        \filldraw (B) circle (1.5pt);
        \coordinate (C) at (1,-.5);
        \filldraw (C) circle (1.5pt);

    \end{tikzpicture}
}

\newcommand{ \loradone}[3] { \begin{tikzpicture}[thick]
        \path [dashed,draw]
        (-1,-.5) node[left]{1} --(0,-.5) node[above]{$#1 $};

        \path [very thick,draw,#2]
        (0,-.5)--(1,-.5) node[above]{$#3 $};

        \path [very thick,draw]
        (0.2,-.5)--(.6,-.5)node[currarrow,scale=1,xscale=1,sloped,pos=.8]{};

        \path [dashed,draw]
        (1,-.5)--(2,-.5) node[right]{$\quad $};

        \path [dashed,draw]
        (-1,-2)  node[left]{2} --(2,-2);

        \path [ draw,gluon]
        (0,-.5) -- (0,-2);

        \path [ draw,gluon]
        (1,-.5) -- (2,-1.5) node[below,right]{$k$};

        \draw[<-,>=stealth] (-0.5,-1) -- (-0.5,-1.6) node[midway, left]
        {$q_{2}$};

        \draw[->,>=stealth] (.8,-.8) -- (1.5,-1.5);

        \coordinate (A) at (0,-.5);
        \filldraw (A) circle (1.5pt);
        \coordinate (B) at (0,-2);
        \filldraw (B) circle (1.5pt);
        \coordinate (C) at (1,-.5);
        \filldraw (C) circle (1.5pt);

    \end{tikzpicture}
}

\def \loradYMTwo { \begin{tikzpicture}[thick]
        \path [dashed,draw]
        (-1,-.5) node[left]{1} --(2,-.5) node[right]{$\quad $};

        \path [dashed,draw]
        (-1,-2) node[left]{2} --(2,-2);

        \path [ draw,gluon]
        (0.5,-.5) -- (0.5,-2);

        \path [ draw,gluon]
        (0.5,-.5) -- (2,-1.5)node[below,right]{$k$};

        \draw[->,>=stealth] (.8,-1) -- (1.4,-1.4);

        \draw[<-,>=stealth] (0.1,-1) -- (0.1,-1.6) node[midway, left]
        {$q_{2}$};
        \coordinate (A) at (.5,-.5);
        \filldraw (A) circle (1.5pt);
        \coordinate (B) at (.5,-2);
        \filldraw (B) circle (1.5pt);
    \end{tikzpicture}
}

\def \loradYMThree { \begin{tikzpicture}[thick]
        \path [dashed,draw]
        (-1,-.5) node[left]{1} --(2,-.5) node[right]{$\quad $};

        \path [dashed,draw]
        (-1,-2) node[left]{2} --(2,-2);

        \path [ draw,gluon]
        (.5,-.5) -- (.5,-2);

        \path [ draw,->]
        (.2,-.7) -- (.2,-1.1)node[midway,left]{$ q_{1}$};

        \path [ draw,->]
        (.2,-1.8) -- (.2,-1.4)node[midway,left]{$ q_{2}$};

        \path [ draw,gluon]
        (.5,-1.3) -- (2,-1.3)node[right]{$k$};

        \path [ draw,->]
        (1,-1.6) -- (1.4,-1.6);

        \coordinate (A) at (.5,-.5);
        \filldraw (A) circle (1.5pt);
        \coordinate (B) at (.5,-2);
        \filldraw (B) circle (1.5pt);
        \coordinate (C) at (0.5,-1.3);
        \filldraw (C) circle (1.5pt);

    \end{tikzpicture}
}

\newcommand{\loradzGR}[3] {
    \begin{tikzpicture}[thick]
        \path [dashed,draw]
        (-1,-.5) node[left]{1} --(0,-.5) node[above]{$#1 $};

        \path [very thick,draw,#2]
        (0,-.5)--(1,-.5) node[above]{$#3 $};

        \path [dashed,draw]
        (1,-.5)--(2,-.5) node[right]{$\quad $};

        \path [dashed,draw]
        (-1,-2) node[left]{2} --(2,-2);

        \path [ draw,double,snake it]
        (0,-.5) -- (0,-2);

        \path [ draw,double,snake it]
        (1,-.5) -- (2,-1.5) node[below,right]{$k$};

        \draw[<-,>=stealth] (-0.5,-1) -- (-0.5,-1.6) node[midway, left]
        {$q_{2}$};

        \draw[->,>=stealth] (.8,-.8) -- (1.5,-1.5);

        \coordinate (A) at (0,-.5);
        \filldraw (A) circle (1.5pt);
        \coordinate (B) at (0,-2);
        \filldraw (B) circle (1.5pt);
        \coordinate (C) at (1,-.5);
        \filldraw (C) circle (1.5pt);

    \end{tikzpicture}
}

\newcommand{ \loradoneGR}[3] { \begin{tikzpicture}[thick]
        \path [dashed,draw]
        (-1,-.5) node[left]{1} --(0,-.5) node[above]{$#1 $};

        \path [very thick,draw,#2]
        (0,-.5)--(1,-.5) node[above]{$#3 $};

        \path [very thick,draw]
        (0.2,-.5)--(.6,-.5)node[currarrow,scale=1,xscale=1,sloped,pos=.8]{};

        \path [dashed,draw]
        (1,-.5)--(2,-.5) node[right]{$\quad $};

        \path [dashed,draw]
        (-1,-2)  node[left]{2} --(2,-2);

        \path [ draw,double,snake it]
        (0,-.5) -- (0,-2);

        \path [ draw,double,snake it]
        (1,-.5) -- (2,-1.5) node[below,right]{$k$};

        \draw[<-,>=stealth] (-0.5,-1) -- (-0.5,-1.6) node[midway, left]
        {$q_{2}$};

        \draw[->,>=stealth] (.8,-.8) -- (1.5,-1.5);

        \coordinate (A) at (0,-.5);
        \filldraw (A) circle (1.5pt);
        \coordinate (B) at (0,-2);
        \filldraw (B) circle (1.5pt);
        \coordinate (C) at (1,-.5);
        \filldraw (C) circle (1.5pt);

    \end{tikzpicture}
}

\def \loradGRTwo { \begin{tikzpicture}[thick]
        \path [dashed,draw]
        (-1,-.5) node[left]{1} --(2,-.5) node[right]{$\quad $};

        \path [dashed,draw]
        (-1,-2) node[left]{2} --(2,-2);

        \path [ draw,double,snake it]
        (0.5,-.5) -- (0.5,-2);

        \path [ draw,double,snake it]
        (0.5,-.5) -- (2,-1.5)node[below,right]{$k$};

        \draw[->,>=stealth] (.8,-1) -- (1.4,-1.4);

        \draw[<-,>=stealth] (0.1,-1) -- (0.1,-1.6) node[midway, left]
        {$q_{2}$};
        \coordinate (A) at (.5,-.5);
        \filldraw (A) circle (1.5pt);
        \coordinate (B) at (.5,-2);
        \filldraw (B) circle (1.5pt);
    \end{tikzpicture}
}

\def \loradGRThree { \begin{tikzpicture}[thick]
        \path [dashed,draw]
        (-1,-.5) node[left]{1} --(2,-.5) node[right]{$\quad $};

        \path [dashed,draw]
        (-1,-2) node[left]{2} --(2,-2);

        \path [ draw,double,snake it]
        (.5,-.5) -- (.5,-2);

        \path [ draw,->]
        (.2,-.7) -- (.2,-1.1)node[midway,left]{$ q_{1}$};

        \path [ draw,->]
        (.2,-1.8) -- (.2,-1.4)node[midway,left]{$ q_{2}$};

        \path [ draw,double,snake it]
        (.5,-1.3) -- (2,-1.3)node[right]{$k$};

        \path [ draw,->]
        (1,-1.6) -- (1.4,-1.6);

        \coordinate (A) at (.5,-.5);
        \filldraw (A) circle (1.5pt);
        \coordinate (B) at (.5,-2);
        \filldraw (B) circle (1.5pt);
        \coordinate (C) at (0.5,-1.3);
        \filldraw (C) circle (1.5pt);

    \end{tikzpicture}
}

\newcommand{\eikonalYMNLOz}[1]{
    \begin{tikzpicture}[thick]
        \begin{feynman}
            \vertex [label=180:$1$] at (0,0) (a0) ;
            \vertex [sdot, right=1 of a0, label=90:$z$] (a1) {};
            \vertex [sdot, right=1 of a1, label=90:$z$] (a2) {};
            \vertex [right=1 of a2] (a3);
            \vertex [below=1.5 of a1] (i1);
            \vertex [sdot] at ($(a1)!1!-30:(i1)$) (b1) {};
            \vertex [left=0.7 of b1] (i2);
            \vertex [label=180:2] at ($(b1)!1!-30:(i2)$) (b0);
            \vertex [right=0.7 of b1] (i3);
            \vertex at ($(b1)!1!-30:(i3)$) (b2);
            \vertex [below=1.5 of a2] (j1);
            \vertex [sdot] at ($(a2)!1!30:(j1)$) (c1) {};
            \vertex [left=0.7 of c1] (j2);
            \vertex [label=180:3] at ($(c1)!1!30:(j2)$) (c0);
            \vertex [right=0.7 of c1] (j3);
            \vertex at ($(c1)!1!30:(j3)$) (c2);
            \diagram*{
                (a0) --[scalar] (a1) --[ultra thick] (a2) --[scalar] (a3),
                (b0) --[scalar] (b1) --[scalar] (b2),
                (c0) --[scalar] (c1) --[scalar] (c2),
                (a1) --[#1] (b1),
                (a1) --[opacity=0, rmomentum'={[compact arrow]$q_2\ \,$}] (b1),
                (a2) --[#1] (c1),
                (a2) --[opacity=0, rmomentum={[compact arrow]$\ \, q_3$}] (c1),
            };
        \end{feynman}
    \end{tikzpicture}
}

\newcommand{\eikonalYMNLOpsi}[3]{
    \begin{tikzpicture}[thick]
        \begin{feynman}
            \vertex [label=180:$1$] at (0,0) (a0) ;
            \vertex [right=1 of a0, sdot, label=90:$#1$] (a1) {};
            \vertex [right=1 of a1, sdot, label=90:$#2$] (a2) {};
            \vertex [right=1 of a2] (a3);
            \vertex [below=1.5 of a1] (i1);
            \vertex [sdot] at ($(a1)!1!-30:(i1)$) (b1) {};
            \vertex [left=0.7 of b1] (i2);
            \vertex [label=180:2] at ($(b1)!1!-30:(i2)$) (b0);
            \vertex [right=0.7 of b1] (i3);
            \vertex at ($(b1)!1!-30:(i3)$) (b2);
            \vertex [below=1.5 of a2] (j1);
            \vertex [sdot] at ($(a2)!1!30:(j1)$) (c1) {};
            \vertex [left=0.7 of c1] (j2);
            \vertex [label=180:3] at ($(c1)!1!30:(j2)$) (c0);
            \vertex [right=0.7 of c1] (j3);
            \vertex at ($(c1)!1!30:(j3)$) (c2);
            \diagram*{
                (a0) --[scalar] (a1) --[ultra thick, fermion] (a2) --[scalar] (a3),
                (b0) --[scalar] (b1) --[scalar] (b2),
                (c0) --[scalar] (c1) --[scalar] (c2),
                (a1) --[#3] (b1),
                (a1) --[opacity=0, rmomentum'={[compact arrow]$q_2\ \,$}] (b1),
                (a2) --[#3] (c1),
                (a2) --[opacity=0, rmomentum={[compact arrow]$\ \, q_3$}] (c1),
            };
        \end{feynman}
    \end{tikzpicture}
}

\newcommand{\eikonalYMNLOcontact}[1]{
    \begin{tikzpicture}[thick]
        \begin{feynman}
            \vertex [label=180:$1$] at (0,0) (a0) ;
            \vertex [sdot, right=1 of a0] (a1) {};
            \vertex [right=1 of a1] (a3);
            \vertex [below=1.5 of a1] (i1);
            \vertex [sdot] at ($(a1)!1!-40:(i1)$) (b1) {};
            \vertex [left=0.7 of b1] (i2);
            \vertex [label=180:2] at ($(b1)!1!-40:(i2)$) (b0);
            \vertex [right=0.7 of b1] (i3);
            \vertex at ($(b1)!1!-40:(i3)$) (b2);
            \vertex [sdot] at ($(a1)!1!40:(i1)$) (c1) {};
            \vertex [left=0.7 of c1] (j2);
            \vertex at ($(c1)!1!40:(j2)$) (c0);
            \vertex [right=0.7 of c1] (j3);
            \vertex [label=0:3] at ($(c1)!1!40:(j3)$) (c2);
            \diagram*{
                (a0) --[scalar] (a1) --[scalar] (a3),
                (b0) --[scalar] (b1) --[scalar] (b2),
                (c0) --[scalar] (c1) --[scalar] (c2),
                (a1) --[#1] (b1),
                (a1) --[opacity=0, rmomentum'={[compact arrow]$q_2\ \,$} ] (b1),
                (a1) --[#1] (c1),
                (a1) --[opacity=0, rmomentum={[compact arrow]$\,\ q_3$} ] (c1),
            };
        \end{feynman}
    \end{tikzpicture}
}

\newcommand{\eikonalYMNLOgluons}[1]{
    \begin{tikzpicture}[thick]
        \begin{feynman}
            \vertex [sdot] (g1) {};
            \vertex [sdot, above=1.3 of g1] (a0) {};
            \vertex [sdot] at ($(g1)!1!120:(a0)$) (b0) {};
            \vertex [sdot] at ($(g1)!1!-120:(a0)$) (c0) {};
            \vertex [left=0.8 of a0, label=180:$1$] (a1);
            \vertex [right=0.8 of a0] (a2);
            \vertex [above=0.8 of b0] (i1);
            \vertex [label=180:$2$] at ($(b0)!1!30:(i1)$) (b1);
            \vertex at ($(b0)!1!180:(b1)$) (b2);
            \vertex [above=0.8 of c0] (i2);
            \vertex [label=0:$3$] at ($(c0)!1!-30:(i2)$) (c1);
            \vertex at ($(c0)!1!180:(c1)$) (c2);
            \diagram*{
                (a1) -- [scalar] (a0) -- [scalar] (a2),
                (b1) -- [scalar] (b0) -- [scalar] (b2),
                (c1) -- [scalar] (c0) -- [scalar] (c2),
                (g1) -- [#1] (a0),
                (g1) -- [#1] (b0),
                (g1) -- [#1] (c0),
                (g1) -- [opacity=0, rmomentum={[arrow distance=7pt, label distance=-2pt, arrow shorten=0.2] $q_1$}] (a0),
                (g1) -- [opacity=0, rmomentum={[arrow distance=7pt, label distance=-2pt, arrow shorten=0.2] $q_2$}] (b0),
                (g1) -- [opacity=0, rmomentum={[arrow distance=7pt, label distance=-2pt, arrow shorten=0.2] $q_3$}] (c0),
            };
        \end{feynman}
    \end{tikzpicture}
}

\title{Classical Double Copy of Spinning Worldline Quantum Field Theory}

\author[a]{Francesco Comberiati}
\author[b]{and Canxin Shi}

\affiliation[a]{Dipartimento di Fisica e Astronomia ``Augusto Righi'', Universit\`a di Bologna \\
    and INFN Sezione di Bologna, via Irnerio 46, I-40126 Bologna, Italy}
\affiliation[b]{Institut f\"ur Physik und IRIS Adlershof,
    Humboldt-Universit\"at zu Berlin, \\
    Zum Gro{\ss}en Windkanal 2, 12489 Berlin, Germany}

\emailAdd{francesco.comberiat2@unibo.it}
\emailAdd{canxin@physik.hu-berlin.de}

\preprint{HU-EP-22/45}

\abstract{
We study the classical double copy of massive spinning objects in the worldline quantum field theories (WQFT) formalism.
We couple the $\mathcal{N}=1$ supersymmetric model to a Yang-Mills background to describe the propagation of a spin-half particle interacting with gluons.
At the classical level, this model captures physical effects up to linear order in spin.
We propose a double copy relation to map the spin tensors to the gravitation side.
Enforcing R-symmetry and supersymmetry (SUSY) on the double copy integrands, 
we find that the gravitational theory is the ${\cal N}=2$ particle coupled to dilaton-gravity (DG).
We check the double copy prescription for the eikonal phase up to next-to-leading order and for radiation at leading order in coupling constants, finding that
the Grassmann nature of the spin tensor in WQFT plays a crucial role in finding full agreement with direct calculation in the $\mathcal{N}=2$ model.
We show how to deform the SUSY charges of the free theory to include DG. 
Since the constraints algebra is first class, the worldline model can be quantized, describing the propagation of a massive vector field coupled to DG, in agreement with the literature.
In addition, we investigate the double copy without preserving SUSY and R-symmetry, finding that the B-field also couples to the~worldline.
}

\begin{document}
\maketitle

\section{Introduction}
General relativity and Yang-Mills theory feature many differences at the quantum level. 
In the weakly interacting case, the quantization of the Einstein-Hilbert action leads to divergent scattering amplitudes in the UV regime, while, on the Yang-Mills side, nice features appear, like renormalizability, leading to 
asymptotic freedom of the theory at high energy. 
Further, the Feynman diagrammatic expansion in the case of gravity is more involved than the Yang-Mills one, given that the diffeomorphism on a Riemannian manifold generates an infinite tower of graviton self-interactions, while on the gauge theory side the diagrammatic expansion is under control even at higher order in perturbation theory, leading to high-precision predictions.
However, Bern, Carrasco, and Johansson (BCJ) showed that, perturbatively, one can relate the scattering amplitudes in quantum gravity to that in gauge theories~\cite{Bern:2008qj, Bern:2010ue}, as the low energy field theory version of the Kawai, Lewellen, and Tye (KLT) relation for open and closed string amplitudes~\cite{kawai1986relation}.
More precisely, the full color-dressed $n$-point tree amplitude in Yang-Mills theory can be organized in terms of trivalent diagrams as
\be 
{\cal A}^{\text{tree}}_{n} = \sum_{i\in \text{trivalent}} \frac{c_{i} n_{i}}{D_{i}},
\ee
where $c_i$ denotes the color factors, while $n_i, D_i$ are the corresponding kinematic numerator and propagator, respectively.
In the case where the color factors satisfy the Jacoby identity
$c_{i} + c_{j} + c_{k}=0$,
one can always arrange the kinematical numerators $n_{i}$ to obey the same algebraic equations $n_{i} + n_{j} + n_{k}=0$. 
This is termed ``color-kinematics duality'' (CKD).
Once such relations are satisfied, the color factors can be replaced with the corresponding kinematic numerators, 
\begin{align}
    {\cal M}^{\text{tree}}_{n} = \sum_{i\in \text{trivalent}} \frac{n_{i} n_{i}}{D_{i}}
\end{align}
yielding an amplitude of a gravitational theory.
This is the so-called double copy relation.
More specifically, the resulting theory is the $N=0$ supergravity (SUGRA or NS-NS gravity), which describes Einstein-Hilbert gravity coupled to the dilaton field $\phi$ and the Kalb-Ramond two-form $B$.
In the Einstein frame, the action\footnote{We use the mostly minus convention for the metric.} in $D$~dimensions is given as \cite{Polchinski:1998rq}:
\be
\label{eq:B-dilaton-gravity}
S_{N=0} = \frac{2}{\kappa^2}\int \dd^D x \sqrt{-g(x)} \left(
-R + \frac{4}{D-2}\partial_\mu \phi \partial^\mu \phi +\frac{1}{12}
e^{-8\phi/(D-2)} H_{\lambda \mu\nu}H^{\lambda \mu\nu}
\right),
\ee
where $H_{\lambda\mu\nu}= \partial_\lambda B_{\mu\nu} +\partial_\mu B_{\nu\lambda} +\partial_\nu B_{\lambda \mu}$ is the field strength related to $B_{\mu\nu}$.


This paved the way for a deeper investigation of this surprising feature shared by perturbative quantum gravity and Yang-Mills.
Such investigations were carried out for a great variety of gauge theories (see \cite{Carrasco:2015iwa, Bern:2019prr, Travaglini:2022uwo, Adamo:2022dcm} for recent reviews).
On the group theoretical side, the color-kinematics duality provides a way to extend gauge invariance of two gauge theories to a diffeomorphism symmetry \cite{Monteiro:2011pc, Anastasiou:2018rdx, Bonezzi:2022yuh}, enjoyed by the gravitational theory.
Furthermore, the color-kinematic relations seem to imply a deeper structure than a simple relation between numerators.
Actually, it has been recently shown how BCJ numerators from amplitudes with one massive scalar line in a gluon background, can be thought as obeying a Hopf algebra \cite{Brandhuber:2021eyq, Brandhuber:2021bsf, Brandhuber:2022enp}.

Most importantly, the double copy provides a framework where calculations in both theories can be carried out by using building blocks arising from the gauge theory, generating results on the gravitational side.
This makes the double copy an efficient tool to generate gravitational amplitudes, in particular for the scattering of massive external lines, which are needed for applications to black hole scattering~\cite{Bern:2019nnu, Bern:2019crd, Bern:2020buy}.
This is important for high-precision predictions~\cite{Buonanno:2022pgc} which will be required by future experiments carried on by the LIGO and VIRGO collaborations \cite{LIGOScientific:2016aoc, LIGOScientific:2017vwq}.
In this direction, the double copy of quantum field theories propagating spinning particles is interesting.
Particularly, Ref.\cite{Maybee:2019jus} showed, by extending the KMOC formalism~\cite{Kosower:2018adc} to include spin, how the double copy of amplitudes with massive vector bosons scattering off scalars, leads to classical amplitudes in a DG background, capturing quadratic effects in the Pauli-Lubanski vector of the spinning particle.
In addition,~\cite{Bautista:2019evw, Johansson:2019dnu} showed that the double copy of scattering amplitudes with fermionic lines leads to vector fields interacting through the dilaton and graviton.
To achieve such a result they employed symmetrization over the external particles polarization indices, inspired by massive spinor helicity formalism \cite{Arkani-Hamed:2017jhn, Ochirov:2018uyq} which makes little group covariance manifest.
These lines of research of course hold at the quantum level, where scattering amplitudes and gauge symmetry can be used as a leading guide to the reconstruction of the double copy theory.

An interesting question how we can apply the double copy relation directly at the classical level.
It is well-known that that some classical solutions of YM theory can be related to general relativity. 
This is first discovered by Monteiro, O’Connell, and White known as the Kerr-Schild double copy~\cite{Monteiro:2014cda}, and then developed to a great extent~\cite{Luna:2018dpt, Chacon:2021hfe, CarrilloGonzalez:2022ggn, Chacon:2021wbr, White:2020sfn, Monteiro:2020plf, Godazgar:2020zbv, Kim:2019jwm, Luna:2016hge, Luna:2016due, Luna:2015paa, Gonzo:2021drq}.
However, there are other approaches that work perturbatively and thus more similar to scattering amplitudes, which can be accomplished at the level of effective action~\cite{Plefka:2018dpa, Plefka:2019hmz} or for classical solutions.
The former turns out to breakdown at next-to-leading order due to the gauge dependence and off-shell nature of the effective action.
The latter is completely on shell and thus more promising.
It is first studied in \cite{Goldberger:2016iau}, and later extended to higher orders~\cite{Shen:2018ebu}, bound states~\cite{Goldberger:2017vcg}, as well as to incorporate spin effects~\cite{Goldberger:2017ogt, Li:2018qap}.
Specifically, in these approaches, the basic idea is to iteratively solve the equations of motion (Wong's equations in the case of a charged particle in YM field~\cite{Wong:1970fu}) to obtain classical observables such as the radiation.
Then, by adopting appropriate replacement rules of the color factors, they recover the corresponding quantities in dilaton-gravity from YM theory.
It should be noted that differing from amplitudes where the locality structure is encoded by the poles in the Feynman propagators, classical integrands do not enjoy such a feature, making it difficult to recast the amplitude in a BCJ form.
A way to tackle such a question was provided by Shen \cite{Shen:2018ebu}, who showed that the double copy at the classical level could be realized in an analogous way to the BCJ one from amplitudes by identifying the so-called ``double copy kernel''.

Recently, Plefka and one of the authors developed an alternative approach to perturbative classical double copy~\cite{Shi:2021qsb}, built upon the worldline quantum field theory (WQFT) formalism developed by Plefka, Mogull, and Steinhoff~\cite{Mogull:2020sak}.
The WQFT is designated to model classical scatterings of compact objects in general relativity, and it has been successfully extended to incorporate spin and finite-size effects~\cite{Jakobsen:2021zvh, Jakobsen:2021lvp, Jakobsen:2021smu, Jakobsen:2022psy, Jakobsen:2022fcj}.
To investigate the classical double copy, the authors in \cite{Shi:2021qsb} generalize the WQFT formalism to the case of massive (colored) point particles coupled to a bi-adjoint scalar, Yang-Mills field, and dilaton-gravity.
The bi-adjoint scalar theory is employed to identify the double copy kernel as in \cite{Shen:2018ebu}.
A double copy prescription for the eikonal phase is proposed and verified by explicit calculations up to subleading order in all three theories.
The advantage of this approach is that classical observables are computed in a diagrammatic way, making it clear to build a connection to scattering amplitudes.

From a more formal perspective, the WQFT provides a direct link between Feynman diagrams and the classical equations of motion for a set of point particles in a gravitational background.
This holds given that such formalism is based on worldline particles, which, once quantized by making use of a path integral \cite{Bastianelli:2005vk, Bastianelli:2005uy, Ahmadiniaz:2015xoa, Bastianelli:2015iba, Bastianelli:2019xhi, Bastianelli:2021rbt} or by using BRST methods \cite{Dai:2008bh, Bonezzi:2018box, Bonezzi:2020jjq}, allows to resum the Feynman diagrams in the quantum theory implemented by the worldline.

Inspired by the above results and, moreover, motivated by the phenomenological applications of the double copy to the classical black hole scattering, we use the quantization procedure with, in addition, the diagrammatic expansion from the WQFT, to study the double copy of spinning worldlines as an extension to \cite{Shi:2021qsb}, so to be able to capture quadratic effects in spin (quadrupole) in the double copy integrands for the binary gravitational radiation and eikonal phase.
This fills a gap in the literature, and, most importantly, provides a general way of dealing with the double copy of WQFTs, when spin is included in the dynamics. 
Indeed, we find that preserving R-symmetry and supersymmetry on the worldline in the double copy procedure, allows us to completely fix both the classical and quantum theory on the gravity side. 
Since SUSY is preserved by our double copy prescription, the constraint algebra stays first class, thus allowing to quantize the worldline particle.
We verify such statements by direct inspection of the double copy integrands for the leading binary radiation and the eikonal phase up to next-to-leading order against the worldline theory we propose as a double copy.

In this article, we start by reviewing the general features of the WQFT and the double copy for scalar particles in section \ref{secReview}.
Then, in section~\ref{secA} we coupled the ${\cal N} = 1$ SUSY model to a Yang-Mills background, evaluating the two-body radiation and three-body eikonal phase.
Next, in section \ref{secDG} we show how to couple the ${\cal N}=2$ worldline particle to the dilaton-gravity background, preserving supersymmetry on the worldline.
Finally, in section \ref{secDC} we show how our double copy procedure from the ${\cal N}=1$ particle, naturally leads to classical integrands generated by the ${\cal N}=2$ model in a DG background, proposing also the quantum field theory related to the double copy of Dirac fermions in four dimensions.

\section{Review of the worldline quantum field theory}\label{secReview}
In this section, we start by briefly reviewing the WQFT formalism and how to extract classical observables by considering the scattering of massive scalar particles in general relativity as an example.
Then, we explain how the double copy relation is realized between worldlines coupled to Yang-Mills theory and to dilaton-gravity, stressing the reduction of the double copy kernel.

\subsection{Basic of worldline quantum field theory}
\label{subsec:WQFTreview}
As introduced in \cite{Mogull:2020sak}, schematically, the action for WQFT consists of a field theory in the bulk and the worldline actions
\begin{align}
    S_{\mathrm{WQFT}} = S_{\mathrm{field}} + \sum S_{\mathrm{worldline}}.
\end{align}
In the case of a massive scalar particle coupled to pure gravity, the worldline action reads
\begin{align}
    \label{eq:LYmeasure}
    S[x_{k}; g] = - \int_{-\infty}^{\infty} \dd\tau \, \frac{m}{2} g_{\mu\nu}
    \left(	\dot{x}_{k}^{\mu}\dot{x}_{k}^{\nu} + a_{k}^{\mu}a_{k}^{\nu} + b_{k}^{\mu}c_{k}^{\nu} \right),
\end{align}
where $x^\mu$ is the spacetime coordinates.
For completion, we have included the bosonic $a^\mu$ and fermionic $b^\mu, c^\mu$ ``Lee-Yang'' ghosts, which arise from the metric-dependent path integral measure~\cite{Bastianelli:1992ct}.
In the bulk,  we have the usual Einstein-Hilbert action,
\begin{align} 
    S_{\textrm{EH}}= -\frac{2}{\kappa^{2}}\int \dd^{4}x \sqrt{-g} R,
\end{align}
where $\kappa = \sqrt{32\pi G_{N}}$ is the coupling constant, with $G_{N}$ being the Newtonian constant of gravitation.

With the action, we can write down the WQFT partition function, which turns out to be a fundamental object in the definition of observables,
\be \label{zdef}
{\cal Z}_{\textrm{WQFT} } = \int {\cal D} g_{\mu\nu} \, e^{i S_{\textrm{EH}}[g]} 
\prod_{k=1}^{n}\int {\cal D} x_{k} \, e^{i S[x_{k}; g]} \,.
\ee
Furthermore, it is noticed that the partition function is related to the eikonal phase via exponentiation~\cite{Mogull:2020sak, Amati:1990xe}
\begin{align}
    \label{eq:eikonaltoZ}
    {\cal Z}_{\textrm{WQFT}} = e^{i \chi}.
\end{align}

To evaluate the path integral \eqref{zdef} one proceeds perturbatively.
Firstly, we expand the metric around the Minkowskian background as
\begin{align}
    g_{\mu\nu} = \eta_{\mu\nu} + \kappa h_{\mu\nu},
\end{align}
where $ h_{\mu\nu}$ is referred to as the graviton.
Additionally, by using translational invariance of the worldline path integral measure, we expand the configuration space variables around a straight line  background as
\be 
\label{eq:xExpansion}
x_{i}^{\mu}(\tau)= b_{i}^{\mu}+ v_{i}^{\mu}\tau + z_{i}^{\mu}(\tau),
\ee
such that the measure in the worldline path integral splits as ${ \cal D }x = {\cal D} z$, leaving a functional integration over the quantum fluctuations $z_{i}(\tau)$ of the $i-$th particle. 
Then, going to energy space for such a fluctuation
\be 
z^{\mu}(\tau) = \int_{-\infty}^{\infty}\frac{\dd \omega}{2\pi} e^{i \omega \tau} z^{\mu}(-\omega):= \int_{\omega}e^{i \omega \tau} z^{\mu}(-\omega),
\ee
one is able to derive the related worldline propagator
\be 
\raisebox{-2mm}{\wprop} = -i\frac{\eta_{\mu\nu}}{\omega^2} \hskip 1cm
\ee
where we do not specify the $i\epsilon$ prescription here since we are only interested in the integrands. 
For more details on this we send the reader to \cite{Mogull:2020sak,Jakobsen:2022psy}.
Then, in order to account interactions, one needs to Fourier expand the graviton field as
\ba
h_{\mu\nu}(x(\tau)) &= \int_{q} \, e^{i q\cdot x(\tau)}h_{\mu\nu}(-q) = \sum_{n=0}^{\infty}\frac{i^{n}}{n!} \int_{q} e^{i q\cdot (b + v\tau)} \left( 	z(\tau)\cdot q	\right)^{n}h_{\mu\nu}(-q)  \\
&= \sum_{n=0}^{\infty}\frac{i^{n}}{n!} \int_{q, \omega_{1}\cdots \omega_{n}}\hskip-.5cm 
e^{i q\cdot b} e^{i (q\cdot v + \sum_{k=1}^{n} \omega_{k})\tau}\left( \prod_{i=1}^{n}
q\cdot z(-\omega_{i})
\right) h_{\mu\nu}(-q) \;.
\ea
Finally, plugging the above expansion in the WQFT action allows one to read out worldline Feynman rules \cite{Mogull:2020sak,Jakobsen:2021zvh,Bastianelli:2021rbt,Shi:2021qsb}, depending on the number of quantum fluctuations $z_i(\omega)$, which can be used to build up a diagrammatic expansion of the partition function in powers of the gravitational coupling constant $\kappa$.

Once having a well-defined partition function, we can identify classical observables as insertion of functions of the worldline variables inside \eqref{zdef}.
This way a classical observable is defined as 
\be \label{ob}
O(b_{i},v_{i}) = \la \hat{O}(\hat{x},\hat{h}_{\mu\nu})\ra =  {\cal Z}^{-1}_{\textrm{WQFT}}
\int {\D h_{\mu\nu}} \, e^{i S_{\textrm{EH} }} 
\left(\prod_{k=1}^{n} \int {\cal D}x_{k} \, e^{i S_{k}}\right) O(x_{i}(\tau),h_{\mu\nu})
\ee
and then, can be computed perturbatively using the Feynman rules derived from the WQFT action as explained above.
The result of this calculation consists in a perturbative expansion of the observable, which is controlled only by the gravitational coupling constant $\kappa$, leading then to what is known as a Post-Minkowskian (PM) expansion.
We can also employ WQFT to compute the gravitational radiation emitted in the process of scattering of the worldlines, since it is going to be used in the next sections.
Following the definition \eqref{ob} we define the gravitational radiation as
\be \label{grad}
-ik^{2} \la h_{\mu\nu}(k)\ra_{\textrm{WQFT}} \Big|_{k^{2}=0}=
{\cal Z}^{-1}_{\textrm{WQFT}}
\int {\D h_{\mu\nu}} \, e^{i S_{\textrm{EH} }} 
\left(\prod_{k=1}^{n} \int {\cal D}x_{k} \, e^{i S_{k}}\right)\left(-i k^{2}h_{\mu\nu}(k)\right)
\ee
where, on the RHS one has to use the on-shell condition on the external graviton i.e. $k^{2}=0$ and contract with a physical polarization tensor for the graviton, so to get a gauge invariant object.

\subsection{WQFT double copy}
\label{subsec:wqftdoublecopy}
The double copy of spinless massive particles in the WQFT formalism was studied by one of the authors and Plefka~\cite{Shi:2021qsb}.
There are two key observations for the double copy in the classical limit of scattering amplitudes.
One is that the propagators have both single and double poles, so the locality structure is unclear.
In order to correctly identify the propagator terms, one can employ the bi-adjoint scalar theory, in which the numerators are trivial in the quantum theory.
In this way, we can obtain the off-diagonal BCJ double copy kernel for classical observables.
The second observation is that at higher order in binary systems, some of the color factors in YM theory will be vanishing due to the anti-symmetry of the structure constant, whereas the corresponding numerators are required for the double copy.
To avoid this problem, we will use as many different worldlines as worldline- field interactions occur.
For example, at leading order of the eikonal phase, two worldlines are sufficient, but at next-to-leading order, we will use three worldlines.

Similar to the case in Einstein gravity, we can consider a massive point charge coupled to Yang-Mills field $A_\mu^a$ in the worldline formalism%
\footnote{Note that our convention is slightly different from \cite{Shi:2021qsb}.
    Specifically, the kinetic terms of the ``color wave function'' are different by a minus sign, leading to a sign change of the color charges.
    Moreover, some quantities are denoted by different symbols.}
\begin{align}
    \label{eq:actionPC}
    S^\mathrm{pc} = \int \! \dd\tau
    \left( - \frac{m}{2} \dot{x}^2
    \!-\! i \bar{c}_\alpha \dot{c}^\alpha
    \!-\! g \dot{x}^\mu A_\mu^a q^a\right),
\end{align}
where $g$ is the coupling constant of YM theory.
We have introduced the bosonic auxiliary ``color wave function'' $\bar{c}_\alpha, c^\alpha$ to carry the color degrees of freedom, with $\alpha$ being the index of the representation of the gauge group.
It is not necessary to choose a specific gauge group, but for convenience, one can use the usual $SU(N)$ and work in the fundamental representation.
The global $SU(N)$ symmetry, $c\to U c$, $\bar{c}\to \bar{c}U^{\dagger}$, gives rise to the conserved current
$q^a = \bar{c}_\alpha (T^a)^\alpha{}_\beta c^\beta$, with $(T^a)^\alpha{}_\beta$,
where $T^{a}$ are the gauge group generators\footnote{In our convention, we choose the commutator of the generators as $[T^a, T^b]= f^{abc} T^c$.} in the fundamental representation,
while we use lowercase Latin letters $a,b,c,\dots$ as the adjoint indices.
This conserved current is interpreted as the color charge of the point particle and is directly related to the classical color in Wong's equations \cite{Wong:1970fu}.
In WQFT, besides the perturbative expansion of coordinate variable $x^\mu$ as in \eqref{eq:xExpansion}, we also expand the color variables around a constant background
\begin{align}
    c^\alpha(\tau) = u^\alpha + \lambda^a(\tau),
\end{align}
and similarly for $\bar{c}_\alpha$.
Consequently, the physical quantities depend on the background color charge
\begin{align}
    C^a = \bar{u}_\alpha (T^a)^\alpha{}_\beta u^\beta.
\end{align}
The double copied worldline theory corresponding to \eqref{eq:actionPC} is a massive scalar particle coupled to dilaton-gravity.
The worldline action simply reads
\begin{align}
    \label{eq:Spmdef}
    S^{\mathrm{pm}} =
    \int \dd \tau
    \left( - \frac{m}{2} e^{2\kappa \phi} g_{\mu\nu} \dot{x}^\mu \dot{x}^\nu \right),
\end{align}
where $\phi$ denotes the dilaton.

The double copy relation between the eikonal of Yang-Mills theory $\chi^{\mathrm{YM}}$ and dilaton-gravity $\chi^{\mathrm{DG}}$ at $\mathrm{N^{(n-1)}LO}$ can be expressed as%
\begin{subequations}
\label{eq:dcEikonal}
\begin{align}
    \label{eq:dcEikonalYM}
    \chi^{\mathrm{YM}}_n =& -(ig)^{2n} \int {\dd \mu_{1,2,...,(n+1)}(0)} \sum_{i,j} \mathcal{C}_i\, \mathcal{K}_{ij}\, \mathcal{N}_j, \\
    \label{eq:dcEikonalDG}
    \chi^{\mathrm{DG}}_n =& -\left( \frac{\kappa}{2} \right)^{2n} \int {\dd \mu_{1,2,...,(n+1)}(0)}\sum_{i,j} \mathcal{N}_i\, \mathcal{K}_{ij}\, \mathcal{N}_j,
\end{align}
\end{subequations}
where $\mathcal{C}_i, \mathcal{N}_j$ are the arrays of color factors and kinematic numerators, respectively, which should be arranged to satisfy color-kinematic duality.
$\mathcal{K}_{ij}$ is the BCJ double copy kernel that are derived from bi-adjoint scalar theory coupled to spinless worldlines, which, for brevity, will not be presented in this article.
We have also defined the integral measure as
\begin{align}
    \label{eq:intmeasure}
    \dd \mu_{1,2,...,n}(k) = \prod_{i=1}^{n}
    \left( \frac{\dd^4 q_i}{(2\pi)^4} e^{i q_i \cdot b_i}
    \deltahat\left(q_i \ccdot p_i \right) \right)
    \deltahat^{(4)} \bigg(\sum_{i=1}^n q_i^\mu - k^\mu\bigg),
\end{align}
with $p_i^\mu = m_i v_i^\mu$ being the kinetic momentum, and $q_i$ is the total outgoing momentum of gluons or gravitons attached to a worldline.

At LO, the double copy structure is simple since there is only one color factor.
We now present the calculation at NLO as an example to show how WQFT double copy works.
The color factors can be arranged as
\begin{align}
    \label{eq:colorNLOold}
    \mathcal{C}_i^{\mathrm{(123)}} =& \big\{
    (C_1\ccdot C_2) (C_1\ccdot C_3), \quad
    (C_1^{ab} C_2^a C_3^b), \quad
    (C_1^{ba} C_2^a C_3^b) \big\} \qquad
    \mathcal{C}_i^{\mathrm{(0)}} = f^{abc} C_1^a C_2^b C_3^c,
\end{align}
where we have used $C_1^{ab} = \bar{u}_\alpha (T^a)^\alpha{}_\beta (T^b)^\beta{}_\gamma  u^\gamma$, and $f^{abc}$ is the structure constant.
Note that there are also $\mathcal{C}_i^{\mathrm{(231)}}, \mathcal{C}_i^{\mathrm{(312)}}$ which can be obtained by simply rotating the indices $(1,2,3)$ in $\mathcal{C}_i^{\mathrm{(123)}}$.
Together, they compose a $10$-dimensional array of color factors.
Due to the commutation relation of the group generators, the color factors satisfy
\begin{align}
    \label{eq:JacobiNLO}
    C_1^{ab} C_2^a C_3^b - C_1^{ba} C_2^a C_3^b = f^{abc} C_1^{c} C_2^a C_3^b.
\end{align}
The BCJ kernel $\mathcal{K}_{ij}$ is block-diagonal,
and the blocks corresponding to the color factors in \eqref{eq:colorNLOold} are
\begin{align}
    \mathcal{K}_{ij}^{\mathrm{(123)}} =&
    \frac{1}{q_2^2 q_3^2} \left(
    \begin{array}{ccccc}
        \frac{q_2 \cdot q_3}{\omega_1^2}  && \frac{1}{\omega_1} && -\frac{1}{\omega_1} \\
        \frac{1}{\omega_1} && 0 && 0 \\
        -\frac{1}{\omega_1} && 0 && 0 \\
    \end{array}
    \right),
    \qquad
    \mathcal{K}_{ij}^{\mathrm{(0)}} = \frac{2}{q_1^2 q_2^2 q_3^2}.
\end{align}
For further convenience, we define%
\footnote{Note that due to different convention, the off-diagonal components of the BCJ kernel are different by a minus sign from \cite{Shi:2021qsb}.}
\begin{align}
    \label{eq:omega_i}
    \omega_1 = p_1 \cdot q_2, \quad
    \omega_2 = p_2 \cdot q_3, \quad
    \omega_3 = p_3 \cdot q_1.
\end{align}
Explicit calculations shows that both the eikonal phases in YM and dilaton-gravity can be expressed in terms of the same numerators
\begin{gather}
    \mathcal{N}_j^{\mathrm{(123)}} =
      \left\{ n_0 \, ,\,  \frac{n_1}{2} \,,  \, \frac{-n_1}{2} \right\}  \qquad
    \mathcal{N}_j^{\mathrm{(0)}} = n_1, \\
    n_0 = p_1\ccdot p_2\, p_1\ccdot p_3 \qquad
    n_1 = q_2 \ccdot p_3\, p_1 \ccdot p_2 - q_3 \ccdot p_2\, p_1 \ccdot p_3 - q_2 \ccdot p_1\, p_2 \ccdot p_3.
\end{gather}
Obviously, they satisfy the color-kinematic duality since $\frac{n_1}{2} - \frac{-n_1}{2} = n_1$.
Similarly, we can obtain the other two blocks of the kernel $\mathcal{K}_{ij}^{\mathrm{(231)}}, \mathcal{K}_{ij}^{\mathrm{(312)}}$ and the corresponding numerators $\mathcal{N}_j^{\mathrm{(231)}}, \mathcal{N}_j^{\mathrm{(312)}}$ by relabeling the labels $(1, 2, 3)$.
Thus the eikonals can be decomposed as
\begin{align}
    \chi^{\mathrm{YM}}_2 =& -g^4 \int \dd \mu_{1,2,3}(0)
      \sum_{i,j} \Big(
      \mathcal{C}_i^{\mathrm{(0)}} \mathcal{K}_{ij}^{\mathrm{(0)}} \mathcal{N}_j^{\mathrm{(0)}}
      + \big(\mathcal{C}_i^{\mathrm{(123)}} \mathcal{K}_{ij}^{\mathrm{(123)}} \mathcal{N}_j^{\mathrm{(123)}}
      + \text{cyclic} \big) \Big)  \\
    \chi^{\mathrm{DG}}_2 =& -\frac{\kappa^4}{16} \int \dd \mu_{1,2,3}(0)
      \sum_{i,j} \Big(
      \mathcal{N}_i^{\mathrm{(0)}} \mathcal{K}_{ij}^{\mathrm{(0)}} \mathcal{N}_j^{\mathrm{(0)}}
      + \big(\mathcal{N}_i^{\mathrm{(123)}} \mathcal{K}_{ij}^{\mathrm{(123)}} \mathcal{N}_j^{\mathrm{(123)}}
      + \text{cyclic} \big) \Big). \nonumber
\end{align}
This prescription agrees with the double copy of scattering amplitudes in scalar QCD, given that the eikonal is directly related to the classical limit of the 6-scalar amplitude~\cite{Plefka:2019wyg}.
This completes the story of the double copy of eikonal phase at NLO in the WQFT formalism.

We note that the 10-dimensional BCJ kernel is actually reducible, as a consequence of the fact that the color factors form a over-completed basis.
Specifically, from $\mathcal{C}_i^{\mathrm{(123)}}$ and $\mathcal{K}_{ij}^{\mathrm{(123)}}$ we see that the contributions to the YM eikonal from the color factors $C_1^{ab} C_2^a C_3^b$ and $C_1^{ba} C_2^a C_3^b$ are different only by a minus sign.
We can thus use the Jacobi identity \eqref{eq:JacobiNLO} to simplify their contributions to $\chi^{\mathrm{YM}}_2$
\begin{align}
    C_1^{ab} C_2^a C_3^b \frac{n_0}{q_2^2 q_3^2 \omega_1}
    + C_1^{ba} C_2^a C_3^b \frac{-n_0}{q_2^2 q_3^2 \omega_1}
    = f^{abc} C_1^{c} C_2^a C_3^b \frac{n_0}{q_2^2 q_3^2 \omega_1}.
\end{align}
This is guaranteed by the fact that $C_1^{ab}, C_1^{ba}$ do not appear in the classical equations of motion, so they must be removed in the final solutions~\cite{Wong:1970fu}.
After the reduction, the YM eikonal can be rewritten as
\begin{align}
    \chi^{\mathrm{YM}}_2 = -g^4 \int \dd \mu_{1,2,3}(0)
      \bigg\{
      &f^{abc} C_1^{c} C_2^a C_3^b 
        \bigg(
          \Big(\frac{n_0}{q_2^2 q_3^2 \omega_1} + \text{cyclic} \Big)
          +\frac{2 n_1}{q_1^2 q_2^2 q_3^2}
        \bigg) \nonumber \\
      +& \bigg(
        (C_1\ccdot C_2) (C_1\ccdot C_3) \Big( 
          \frac{q_2 \ccdot q_3\, n_0}{q_2^2 q_3^2 \omega_1^2}
          + \frac{n_1}{q_1^2 q_2^2 \omega_1} \Big)
        + \text{cyclic} \bigg).
      \bigg\}
\end{align}
We can reduce the full 10-dimensional BCJ double copy kernel to $4$ dimensions
\begin{align}
    \label{eq:BCJkernelNLO}
    \mathcal{K}_{ij} =
    \left(
    \begin{array}{cccc}
        \frac{q_2 \cdot q_3}{q_2^2 q_3^2 \omega_1^2} & 0 & 0 & \frac{1}{q_2^2 q_3^2 \omega_1} \\
        0 & \frac{q_1 \cdot q_3}{q_1^2 q_3^2 \omega_2^2} & 0 & \frac{1}{q_1^2 q_3^2 \omega_2} \\
        0 & 0 & \frac{q_1 \cdot q_2}{q_1^2 q_2^2 \omega_3^2} & \frac{1}{q_1^2 q_2^2 \omega_3} \\
         \frac{1}{q_2^2 q_3^2 \omega_1} & \frac{1}{q_1^2 q_3^2 \omega_2} & \frac{1}{q_1^2 q_2^2 \omega_3} & \frac{2}{q_1^2 q_2^2 q_3^2}
    \end{array}
    \right),
\end{align}
The associated arrays of color factors and numerators are 
\begin{gather}
    \label{eq:colorNLOnew}
    \mathcal{C}_i = \big\{
      (C_1\ccdot C_2) (C_1\ccdot C_3), \quad
      (C_1\ccdot C_2) (C_2\ccdot C_3), \quad
      (C_1\ccdot C_3) (C_2\ccdot C_3), \quad
      f^{abc} C_1^a C_2^b C_3^c \big\} \\
    \label{eq:numeratorsNLOnew}
    \mathcal{N}_j = \big\{
      n_0, \quad n_0^\prime, \quad n_0^{\prime\prime}, \quad n_1 \big\},
\end{gather}
where $n_0^\prime = p_1\ccdot p_2\, p_2\ccdot p_3$ and $n_0^{\prime\prime} = p_1\ccdot p_3\, p_2\ccdot p_3$ are obtained by relabeling $(1,2,3)$ in $n_0$.

In this new basis of the color factors, we no longer explicitly have the Jacobi identities, whereas the color-kinematics duality is hidden in that such a decomposition is possible.
The number of independent color factors also agrees with that of the $6$-quark amplitude in QCD in the Melia basis~\cite{Johansson:2015oia}.
It is straightforward to check that the double copy relation \eqref{eq:dcEikonal} still works.

The advantage of the reduced BCJ kernel is not only the lower dimension, but, more importantly, that it is invertible.
Therefore, one can easily do the KLT-like double copy by inverting the BCJ kernel, as we will show in subsection~\ref{subsec:dcEikonal}.
However, this is not true for kernel of the radiation, which is degenerate.
As we will see in subsection~\ref{subsec:YMradiation}, this has to do with the Ward identity.

\section{QCD on the worldline}\label{secA}
In this section we are going to build up a worldline model which in first quantization is able to propagate a massive Dirac fermion coupled to a non-Abelian gauge background, which is exactly what we need to capture linear terms in spin in our classical applications, as first showed on the amplitude side in \cite{Maybee:2019jus}. As we will see, with the gauge fixing choice done in
\cite{Jakobsen:2021zvh}, 
our action will reduce to the one in \cite{Goldberger:2017ogt}, thus showing the equivalence between the two formulations and giving new insight on the relation between double copy and supersymmetry on the worldline, when the latter is preserved.
When switching off all the interactions such a model is known as the ${\cal N}=1$ SUSY particle on the worldline and was first formulated in \cite{Brink:1976sz}, then coupled to gravity on the torus in \cite{Alvarez-Gaume:1983ihn} for computing gravitational anomalies, and recently used to compute Feynman diagrams in QED in \cite{Ahmadiniaz:2021gsd} where a path integral on the line has been implemented to accomplish such a task.

\subsection{Coupling to Yang-Mills}
Let us start by building up a worldline model that describes the propagation of a spin-half particle in a Yang-Mills background.
We will then explain how to perform classical calculations.
The model can be formulated by introducing a set of real Grassmann variables $\psi_{M} = (\psi_\mu, \theta)$ alongside the usual bosonic variables $x^M =(x^\mu,x^5), P_{_M} = (P_\mu, P_5)$, with $\mu$ being a Lorentz index while the raising and lowering procedure is done by $\eta_{_{MN}}=\textrm{diag} \left( \eta_{\mu\nu}, -1 \right)$.
As we will see, the Grassmann variables will take care of the spinning degrees of freedom of the particle propagated in first quantization by the worldline model, while the auxiliary fifth component has been introduced so to be able to give a mass to such a particle by the Kaluza-Klein dimensional reduction.
We consider the following phase space action
\be \label{sph1}
S_{\mathrm{ph}} = -\int_0^1 \dd\tau \left(
\dot{x}^M P_M +\frac{i}{2}\psi_M \dot{\psi}^M +i \bar{c}_\alpha \dot{c}^\alpha -e H -i\chi Q
\right),
\ee
where we gauge the reparametrization invariance through the gauge field $e$, known as the einbein, and its generator $H$ representing the point particle Hamiltonian.
Further, we also gauge the worldline supersymmetry through the Grassmann-valued gauge field $\chi$, known as the gravitino, and the correspondent generator $Q$ which is the SUSY charge, i.e., the conserved charge under supersymmetry transformation of the worldline variables.

One could even gauge a $U(1)$ worldline symmetry on the color sector, including a Chern-Simmons like coupling term $a (\bar{c} \cdot c - s)$ in the above action, where $a$ is a worldline gauge field and $s$ is an integer parameter.
Particularly, as showed in \cite{Bastianelli:2015iba, Ahmadiniaz:2015xoa} and extended in \cite{Bastianelli:2021rbt} for the case of the bi-adjoint scalar, this coupling is needed to project on a specific field that has $s$ indices of the fundamental representation of the color group.
However, by explicitly performing calculations, as also done in \cite{Shi:2021qsb}, we see that such a gauging is not necessary for classical applications, meaning that propagating all of the color representations in the WQFT consistently implements the classical limit for the color degrees of freedom.

The phase space action \eqref{sph1} allows us to read out Poisson brackets between canonical coordinates, namely
\be  \label{pb}
\lbrace x^\mu, P_\nu	\rbrace=\delta^\mu{}_\nu, \hskip.3cm
\lbrace \psi^\mu,\psi_\nu   \rbrace =-i\delta^\mu{}_\nu, \hskip.3cm 
\lbrace \theta,\theta   \rbrace=i, \hskip.3cm 
\lbrace c^\alpha,\bar{c}_\beta 		\rbrace =-i\delta^\alpha{}_\beta
\ee
with all of the remaining Poisson brackets vanishing.
The Kaluza-Klein reduction here is simply implemented by fixing $p_5 = m$ and gauge away $x_5$ since, even after coupling to background fields, it will appear as a total derivative.

In order to couple to a Yang-Mills background we define the SUSY charge as
\be \label{susy-color}
Q= \psi^\mu\left(P_\mu-gA_\mu^a q_a	\right) - m\, \theta,
\ee
such that, by using the SUSY algebra we can fix the point particle Hamiltonian
\be \label{h}
\lbrace Q,Q\rbrace =-2i H =  -2i\left( \frac{1}{2}\left(\pi^2-m^2 \right) 
-\frac{g}{2} S^{\mu\nu} F_{\mu\nu}^a q_a \right),
\ee
where $\pi_\mu = P_\mu- g A_\mu^a q_a$ is the covariant momentum of the particle while, $F_{\mu\nu}^a = 2\partial_{[\mu}A^a_{\nu]}-ig f^{abc}A_\mu^b A_\nu^c$ is the Yang-Mills field strength.
We have also defined the spin tensor as
\begin{align}
     S^{\mu\nu} = -i \psi^\mu \psi^\nu,
\end{align}
which is the conserved current under the Lorentz symmetry on the fermionic sector of the worldline i.e. $\psi \to e^{\omega_{\mu\nu}J^{\mu\nu}}\psi$ with $\omega_{\mu\nu}$ and $J^{\mu\nu}$ being respectively the Lorentz group parameters and the generators in the fundamental representation.
One can check that the Possion bracket $\{S^{\mu\nu}, S^{\rho\sigma}\}$ agrees with the Lorentz algebra.

Let us now turn to the quantization of the model, so to explicitly see which is the particle propagated by the worldline in first quantization. 
In order to recover unitarity at the quantum level, we need to check that the constraint algebra is first class, namely that
\be
\lbrace Q,Q\rbrace = -2i H, \hskip.4cm \lbrace Q, H\rbrace = 0,
\ee
which is straightforward in such a case and holds by construction.
Then we move to the quantization
by promoting the Poisson brackets to graded commutators as $\{\bullet,\, \bullet\} \to -i[\bullet,\, \bullet\rbrace$. This allows us to get the quantum algebra
\be
[\hat{x}^{\mu},\hat{P}_{\nu}] = i \delta^{\mu}_{\nu}, \hskip.3cm
\lbrace \hat{\psi}^{\mu},\hat{\psi}^{\nu}\rbrace =  \eta^{\mu\nu}, \hskip.3cm
\lbrace \hat{\theta},\hat{\theta} \rbrace =  -1, \hskip.3cm   [\hat{c}^{\alpha },\hat{c}^{\dagger}_{ \beta }] =  \delta^{\alpha}_{\beta}
\,.
\ee
We represent $\hat{P}_\mu = -i \partial_\mu$ and $\hat{x}_\mu$ acting as a multiplication on states, further, once rescaling $\psi^\mu \to \frac{1}{\sqrt{2}}\psi^{\mu},\theta \to \frac{i}{\sqrt{2}}\theta $ we can realize the Grassmann variables as $\hat{\psi}^\mu = \gamma^\mu$, $\hat{\theta} = i\gamma_5$, generating then the Clifford algebra $\lbrace \gamma^\mu,\gamma^\nu\rbrace = 2\eta^{\mu\nu}$ alongside with $\lbrace \gamma^{\mu},\gamma_{5}\rbrace = 0$.
Instead, the color variables can be naturally realized at the quantum level as $\hat c^{\dagger} = {\bar c}$, $\hat{c}= \partial/\partial \bar c$, thus being creation and annihilation operators for a set of oscillators.
Then, one can use the coherent state basis related to such operators, to expand a generic wave function propagated by the worldline as
\be
\Phi(x,\bar{c}) = \sum_{n=0}^{\infty}\frac{1}{n!} \Phi_{\alpha_{1} \alpha_{2} \cdots \alpha_{n}}(x) \bar{c}^{\alpha_{1} }
\bar{c}^{\alpha_{2} }\cdots \bar{c}^{\alpha_{n} }\;.
\ee
Recalling the color variables are charged under the $SU(N)$ global symmetry, as said previously, implies that the worldline particle is propagating totally symmetric tensor products of the fundamental representation.
To project out on a $s$-tensor product, we use the gauged $U(1)$ constraint\footnote{We use normal ordering to solve the ordering ambiguities arising when writing quantum constraints from classical ones.} at the quantum level
\be
\left( \bar{c}^{\alpha}\frac{\partial}{\partial \bar{c}^{\alpha}} -s\right) \Phi(x,\bar{c})=0,
\ee
uniquely selecting the component of the wave function with $s$ indices in the fundamental representation of the color group
\be
\Phi_{s}(x,\bar{c}) = \frac{1}{s!}\Phi_{\alpha_{1} \alpha_{2} \cdots \alpha_{s}}(x) \bar{c}^{\alpha_{1} }
\bar{c}^{\alpha_{2} }\cdots \bar{c}^{\alpha_{s} }\,.
\ee
Then, we impose the equations of motion of the worldline gauge fields $(e, \chi)$ as operator constraints on the above wave function.
In particular, the equation of motion for the gravitino delivers 
\be
\frac{\delta S_{\mathrm{ph}}}{\delta \chi } = 0 \quad \to \quad 
\hat{Q}\Phi_{s}(x) = 0 \quad \to \quad
\left( \gamma_5 \slashed{D} +m 	\right)\Phi_{s}(x,\bar{c}) =0
\ee  
with the gauge covariant derivative $D_\mu= \partial_\mu -ig A_\mu^a q_a$ acting in the $s$ representation.
Particularly, choosing $s=1$, allows us to project on a colored Dirac Fermion, as stressed by the above equation of motion.
The latter can be recast as the standard Dirac equation from textbooks, by performing a change of basis in the spinor space, keeping invariant the Clifford algebra, namely defining $i \tilde{\gamma}_\mu =- \gamma_5 \gamma_\mu$.

This analysis reveals that the worldline supersymmetric ${\cal N}=1$ model, with the SUSY charge deformation
\eqref{susy-color},
allows to propagate a colored Dirac spinor in the fundamental representation of $SU(N)$. Thus, the path integral quantization allows to resum Feynman diagrams with two external spin-half particles in a non-Abelian gauge background.
Noticeably, recalling that the expansion of classical observables at linear order in the spin tensor is reproduced by scattering amplitudes of Dirac fermions \cite{Maybee:2019jus,Guevara:2019fsj,Bautista:2019evw,Guevara:2017csg}, the above analysis makes it clear why one must use the ${\cal N}=1$ SUSY model to compute classical observables at linear order in spin.

After this digression we can come back to the action principle needed for our classical applications.
Now we can plug the Hamiltonian and the supercharge \eqref{h} inside the phase space action \eqref{sph1}. 
Then, eliminating the momentum, by using its equation of motion
\be 
\frac{\delta S_{\mathrm{ph}}}{\delta P_\mu} =0 \quad \rightarrow \quad
P_\mu = e^{-1}\left(\dot{x}_\mu +e g A_\mu^a q_a-i\chi \psi_\mu		\right)
\ee 
and plugging it back in the phase space action \eqref{sph1}, we can write down the worldline action for a color-charged point particle in configuration space as follows 
\ba
S_{\mathrm{pc}} = \int_0^1 \dd\tau \Big(&-\frac{1}{2}e^{-1} \left(\dot{x}^2 +e^2m^2\right)-\frac{i}{2}\psi\cdot \dot{\psi}+\frac{i}{2}\theta \dot{\theta}-i \bar{c}\cdot \dot{c}\\
& -g \dot{x}\cdot A^a q_a-\frac{eg}{2}S^{\mu\nu}F_{\mu\nu}^a q_a+ie^{-1}\chi  \dot{x}\cdot \psi + i m\chi \theta 
\Big).
\ea
Then, following \cite{Jakobsen:2021zvh}, imposing the constraints $\theta=0,\chi =0$ yields $\dot{x}\cdot \psi =0$, which implies the spin supplementary condition (SSC), $\dot{x}_\mu S^{\mu\nu} = 0$.
Furthermore, we gauge fix $e= 1/m$, then we change the integration boundaries to $(-\infty, \infty)$ as a consequence of the LSZ reduction procedure on the external legs \cite{Mogull:2020sak,Bastianelli:2021rbt}.
Further, we rescale $\tau\to m\tau$ such that the worldline action, ready to be used to perform classical calculations, reads as
\be \label{actionWQFT}
S_{\mathrm{pc}} =- \int_{-\infty}^\infty \dd\tau \left(
\frac{1}{2} \dot{x}^2
+\frac{i}{2} \psi\cdot \dot{\psi} +i \bar{c}\cdot \dot{c} +g\dot{x}\cdot A^a q_a +\frac{g}{2} S^{\mu\nu}F_{\mu\nu}^a q_a
\right),
\ee
where we have dropped an unimportant constant mass term proportional to $m^2$.
We note that due to the rescaling of the integration variable, $\dot{x}^\mu$ has the dimension of momentum, so that in perturbative calculations of the WQFT, it is easier to make contact with integrands arising from the classical limit of scattering amplitudes.

Before ending this subsection, let us give some comments on the supersymmetry on the worldline and the double copy.
The action \eqref{actionWQFT} is in agreement with the one used by Goldberger, Li and Prabhu in \cite{Goldberger:2017ogt} to study the double copy of classical spinning particles at linear order in spin. 
In such a case they showed that, in order to have a double copy radiation satisfying the linearized Ward identities for gravity, one must fix the coupling of the Pauli interaction $S_{\mu\nu}F^{\mu\nu}$ to be minus one half (in such conventions). 
However, in our consideration we get such a coupling, very surprisingly, completely by the SUSY algebra on the worldline, thus implying that SUSY, at linear level in spin, allows for a consistent double copy of spinning worldlines!
This may also be tracked back to the fact that the constraint algebra is of first class, thus the model can be consistently quantized, propagating a Dirac spinor coupled to Yang-Mills, which leads to consistent double copy of amplitudes in four dimensions at the quantum level as shown in \cite{Bautista:2019evw, Johansson:2019dnu}.

\subsection{Feynman rules}
\label{subsec:YMFeyn}
Let us now move to the derivation of the vertices arising from the above action.
First, thanks to the rescaling performed in the previous section, we can expand the position space variables in a straight-line background as $x^\mu(\tau)= b^\mu + p^\mu \tau + z^\mu(\tau)$ so to write Feynman rules in terms of the momentum $p^\mu$ of the worldline.
Then, for the spin and color wave function we use the background expansions
\be \label{be-real-psi}
 \psi^\mu(\tau)= \zeta^\mu + \Psi^\mu(\tau), \hskip.4cm
c_a(\tau) = u_a + \lambda_a(\tau) 
\ee
with the expansion for $\bar{c}^a(\tau)$ obtained by complex conjugation.
Particularly these expansions allows us to identify the classical values of the spin tensor and the color charge of the worldline particle, namely ${\cal S}_{\mu\nu} = -i \zeta_\mu \zeta_\nu$ and $C^a = \bar{u}_\alpha (T^a)^\alpha{}_\beta \, u^\beta$ respectively.
In order to derive the Feynman rules, we expand all of the quantum fluctuations in energy space
\be 
z^\mu(\tau) = \int_\omega \, e^{i \omega \tau} z^\mu(-\omega),\hskip.3cm 
\Psi^\mu(\tau) = \int_\omega e^{i \omega \tau} \Psi^\mu(-\omega), \hskip.3cm 
\lambda_a(\tau)=\int_\omega \, e^{i \omega \tau} \lambda_a(-\omega),
\ee
where the expansion for $\bar{\lambda}$ is obtained by simply taking the complex conjugate of $\lambda$.
This way, we can write down the worldline propagators\footnote{Since we are only interested in writing down classical integrands, we avoid picking a particular $i\epsilon$ prescription on our worldline and bulk propagators, more details on this can be found in \cite{Jakobsen:2022psy}.} related to each of the quantum fluctuations alongside with the gluon propagator
\ba
\raisebox{-2mm}{\wprop} &= -i\frac{\eta_{\mu\nu}}{\omega^2} \hskip 1cm
\raisebox{-2mm}{\gluprop} = -\frac{i}{q^2}\eta_{\mu\nu} \delta^{a b} \\
\raisebox{-2mm}{\sprop} &= i\frac{\eta_{\mu\nu}}{\omega} \hskip1.2cm
\raisebox{-2mm}{\cprop} = \frac{i}{\omega}\delta_\alpha{}^\beta  \;.
\ea
Then, we move to interactions by writing the Yang-Mills field in momentum space, further expanding for the quantum fluctuations in energy space
\ba
A_\mu^a(x(\tau)) &= \int_q \, e^{iq\cdot x(\tau)}\, A_\mu^a(-q) = \sum_{n=0}^\infty \frac{i^n}{n!}\int_q \,e^{i q\cdot(b + p\tau)} \left(q\cdot z(\tau)	\right)^n A_{\mu}^{a}(-q)\\
&=\sum_{n=0}^{\infty}\frac{i^{n}}{n!} \int_{q, \omega_{1}\cdots \omega_{n}}\hskip-.5cm 
e^{i q\cdot b} e^{i (q\cdot p + \sum_{k=1}^{n} \omega_{k})\tau}\left( \prod_{i=1}^{n}
q\cdot z(-\omega_{i})
\right) A_\mu^a(-q)
\ea
such that, plugging the above expansion in the interactions from \eqref{actionWQFT} allows us to write down the worldline Feynman rules.

Here we start by listing the lowest order rules,
\ba \label{LORule}
&\raisebox{-10mm}{\vglu }=- ig C^a e^{iq\cdot b}\deltahat(q\cdot p) \left(	p^\mu + i (q\cdot \s)^ \mu	\right)   \\
&\raisebox{-10mm}{\vKinglu }= g C^a  e^{iq\cdot b}\deltahat(q\cdot p+\omega )
\left(
p^\mu q^\alpha + \omega \eta^{\mu\alpha} +i (q\cdot \s)^\mu q^\alpha 
\right) 
\ea
where $(q\cdot \s)^\mu = q_\nu \s^{\nu\mu}$. One can notice that, up to on-shell terms, the first vertex above corresponds to the classical piece in the three point amplitude for two spin half fermions emitting a gluon, as expected by the previous quantization procedure.
Next we move to the vertices propagating fluctuations coming from the Grassmann and color variables respectively
\ba
\raisebox{-10mm}{\vSpinglu} &= -2i g C^a e^{i q\cdot b}\deltahat(q\cdot p + \omega) \zeta^\rho q_{[\rho} \eta_{\nu]\mu} \\
\raisebox{-10mm}{\vColglu} &= -i g e^{iq\cdot b}\deltahat(q\cdot p+\omega ) \left(
p^\mu +i (q\cdot \s)^\mu 
\right) \bar{u}_\beta  (T^a)^\beta{}_\alpha \\
\raisebox{-10mm}{\vbarColglu} &= -i g e^{iq\cdot b}\deltahat(q\cdot p-\omega ) \left(
p^\mu +i (q\cdot \s)^\mu 
\right) (T^a)^\alpha{}_\beta u^\beta
\ea
In calculation of Feynman diagrams, we follow the arrow of $(\Psi, \lambda)$ propagators to combine the vertices.
we also get a vertex with the emission of two gluons from the worldline, namely
\be
\raisebox{-7mm}{\vtwoglu} =- g^2 \, f^{bcd}C^d e^{i (q_1+q_2)\cdot b}\deltahat((q_1+q_2)\cdot p)\s^{\mu\nu},
\ee
as a direct consequence of SUSY on the worldline.

\subsection{Yang-Mills eikonal phase}
As introduced in subsection~\ref{subsec:WQFTreview}, the WQFT partition function captures the information needed to compute classical observables.
Specifically, in the $\mathcal{N} = 1$ model, it gives predictions up to linear terms in spin, corresponding to the scatterings of spin-$\half$ charged particles.
For a model with $n$ worldlines described by the action \eqref{actionWQFT}, the partition function reads as follow
\be
{\cal Z}_{\textrm{YM}} = \int {\cal D} A  \, e^{i S_{\textrm{YM}}}\, \int \prod_{k=1}^n \, {\cal D}X_k \, e^{i S[X_k;A]},
\ee
where $S_{\textrm{YM}}$ is the Yang-Mills action in the Feynman gauge, and we collect all of the worldline variables in $X =(x,\psi,c,\bar{c})$.
The eikonal phase is related to the partition function in the same way as \eqref{eq:eikonaltoZ}.
With the background expansion of the worldline variables and the Feynman rules provided in the previous subsection, the eikonal can be calculated perturbatively.
It corresponds to the classical limit of the scattering amplitude with massive external particles, thus is suitable for the double copy.

\paragraph{Leading order eikonal:}
At leading order (LO), only one diagram contributes to the eikonal.
Using the vertex in \eqref{LORule}, we obtain
\begin{align}
\label{eq:YMeikonalLO}
    i \chi^{\mathrm{YM}}_1 = \raisebox{-9mm}{\loeikYM} =
        i g^{2} C_{1}\cdot C_{2} \int &\frac{\dd \mu_{1,2}(0)}{q_2^2}
        \big(\! \gamma
        - i q_2\cdot \s_1 \cdot p_{2} \nonumber \\[-6mm]
        & + i q_2 \cdot \s_2 \cdot p_1
        - q_2 \cdot \s_1 \cdot \s_2 \cdot q_2 \big),
\end{align}
where we have defined $\gamma = p_1\cdot p_2$, and used the notation
$a \cdot \s_i \cdot \s_j \cdot b = a_\mu (\s_i)^{\mu\nu} (\s_j)_{\nu\rho} b^\rho$ for arbitrary vectors $a^\mu, b^\mu$.
We can recover the scalar WQFT result by simply setting the spin to zero, and it agrees with the result in \cite{Shi:2021qsb}.
In the spirit of the double copy, we identify the color factor, the double copy kernel and the numerator in the form \eqref{eq:dcEikonal} as
\begin{gather}
    \mathcal{C} = C_1 \cdot C_2, \qquad
    \mathcal{K} = \frac{1}{q_2^2}, \\
    \mathcal{N} = \gamma
      - i \big( q_2 \cdot \s_1 \cdot p_2
        - q_2 \cdot \s_2 \cdot p_1 \big)
      - q_2 \cdot \s_1 \cdot \s_2 \cdot q_2.
\end{gather}

\paragraph{Next-to-leading order eikonal:}
For next-to-leading order (NLO) eikonal of two worldlines, some of the color factors will be vanishing due to the anti-symmetry of structure constant, whereas the corresponding numerators are needed in the double copy.
To circumvent this problem, we will consider three bodies at NLO.
One can easily retrieve the binary system by identifying two of the three worldlines.

\begin{figure}[t]
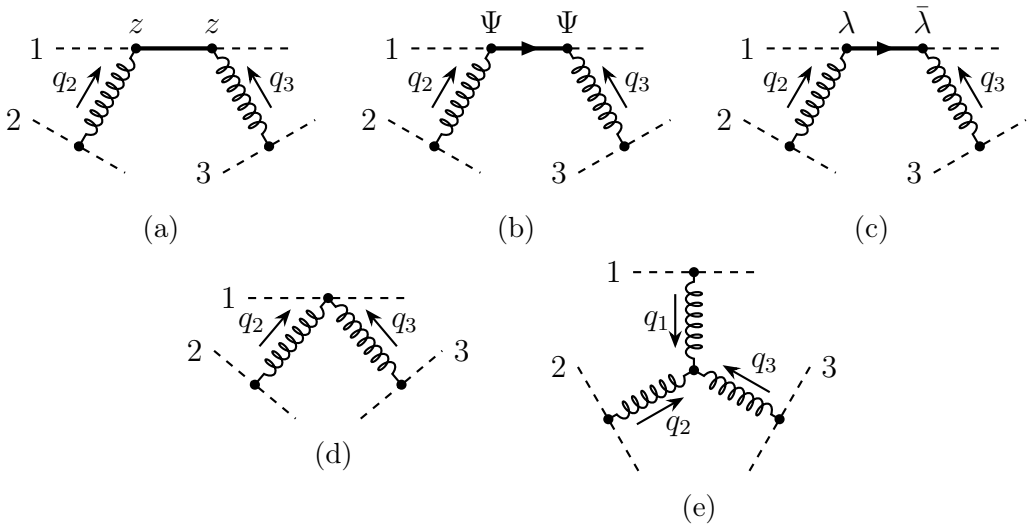

    \centering
    \begin{subfigure}[b]{0.3\textwidth}
        \centering
        \eikonalYMNLOz{gluon}
        \caption{}
        \label{fig:eikonalNLOz}
    \end{subfigure}
    \begin{subfigure}[b]{0.3\textwidth}
        \centering
        \eikonalYMNLOpsi{\Psi}{\Psi}{gluon}
        \caption{}
        \label{fig:eikonalNLOpsi}
    \end{subfigure}
    \begin{subfigure}[b]{0.3\textwidth}
        \centering
        \eikonalYMNLOpsi{\lambda}{\bar{\lambda}}{gluon}
        \caption{}
        \label{fig:eikonalNLOc}
    \end{subfigure}
    \\
    \begin{subfigure}[b]{0.3\textwidth}
        \centering
        \eikonalYMNLOcontact{gluon}
        \caption{}
        \label{fig:eikonalNLOcontact}
    \end{subfigure}
    \raisebox{-7mm}{
    \begin{subfigure}[b]{0.3\textwidth}
        \centering
        \eikonalYMNLOgluons{gluon}
        \caption{}
        \label{fig:eikonalNLO3gluon}
    \end{subfigure}}
    \caption{The diagrams for the NLO eikonal with three worldlines.
        For diagrams involving the propagators of $\Psi$ (\ref{fig:eikonalNLOpsi}) and $\lambda$ (\ref{fig:eikonalNLOc}), we also need to include the crossed diagrams that can be obtained by simply reversing the arrows on the worldline.
        We only display diagrams with worldline propagator and contact interaction of particle~1.
    }
    \label{fig:eikonalNLO}
\end{figure}
The diagrams with three worldlines are collected in Fig.\ref{fig:eikonalNLO}.
It is then straightforward to compute the contributions to the eikonal phase with the Feynman rules given in \ref{subsec:YMFeyn}.
The full result can be formally expressed as
\begin{align}
    \label{eq:YMeikonalNLO}
    \chi^{\mathrm{YM}}_2 = -\! g^4 \int \dd\mu_{1,2,3}(0)
      &\Big( f^{abc} C_1^a C_2^b C_3^c\, \mathcal{A}^{(0)}
      + \left( C_1\ccdot C_2\, C_1 \ccdot C_3 \,
        \mathcal{A}^{(1)}
        + \text{cyclic} \right) \Big),
\end{align}
where $\mathcal{A}^{(i)}$ are gauge-invariant ``partial eikonals'' akin to color-ordered amplitudes at the quantum level.
Explicitly, they can be written as
\begin{gather}
    \mathcal{A}^{(0)} = \Big(
    \frac{2}{q_1^2 q_2^2 q_3^2} n^{(0)}
    + \frac{1}{q_2^2 q_3^2 \omega_1} n^{(123)}
    + \frac{1}{q_1^2 q_3^2 \omega_2} n^{(231)}
    + \frac{1}{q_1^2 q_2^2 \omega_3} n^{(312)}
    \bigg) \\
    \mathcal{A}^{(1)} = \Big( \frac{1}{q_2^2 q_3^2 \omega_1} n^{(0)}
    + \frac{q_2 \ccdot q_3}{q_2^2 q_3^2 \omega_1^2} n^{(123)} \Big)
\end{gather}
where $\omega_i$ is defined in the same way as the spinless case~\eqref{eq:omega_i}, and the numerators read
\begin{align}
    \label{eq:NLOnumerator123}
    n^{(123)}\!=&
    \half p_1\cdot p_2\, p_1\cdot p_3
    +i \big( p_1\cdot p_3 \left(q_2\cdot \s_2\cdot p_1
    -q_2\cdot \s_1\cdot p_2\right)
    + \half q_3\cdot p_1\, p_2\cdot \s_1\cdot p_3 \big)
    \nonumber \\
    & + q_2\ccdot \s_2\ccdot p_1 \left( q_3\ccdot \s_1\ccdot p_3 -\half q_3\ccdot \s_3\ccdot p_1 \right)
    -p_1\ccdot p_3\, q_2\ccdot \s_1\ccdot \s_2\ccdot q_2
    -q_3\ccdot p_1\, q_2\ccdot \s_2\ccdot \s_1\ccdot p_3
    \nonumber \\
    &+i \big(\half q_3\cdot p_1\, q_2\cdot \s_2\cdot \s_1\cdot \s_3\cdot q_3
    -q_2\cdot \s_1\cdot \s_2\cdot q_2\, q_3\cdot \s_3\cdot p_1 \big)
    + (2 \leftrightarrow 3)\\
    n^{(0)}\! =& p_1\cdot p_2\, q_2\cdot p_3 
    -i \left(
    p_2\cdot p_3\, q_3\cdot \s_1\cdot q_1
    +q_1\cdot p_3\, q_1\cdot \s_1\cdot p_2
    +q_3\cdot p_2\, q_1\cdot \s_1\cdot p_3 \right)
    \nonumber \\
    &+q_2\cdot p_3\, q_1\cdot \s_1\cdot \s_2\cdot q_2
    +q_2\cdot \s_1\cdot q_3\, q_2\cdot \s_2\cdot p_3
    +q_2\cdot \s_2\cdot q_3\, q_1\cdot \s_1\cdot p_3
    \nonumber \\
    &-i q_1\cdot \s_3\cdot q_2\, q_1\cdot \s_1\cdot \s_2\cdot q_2 
    + \text{cyclic},
\end{align}
and $n^{(231)}$ and $n^{(312)}$ are obtained by relabeling $(1, 2, 3)$ in \eqref{eq:NLOnumerator123}.
We stress that the $\mathcal{N}=1$ model captures only linear terms in spin, so the numerators are truncated up to linear order in each of the spin tensor.

In \eqref{eq:YMeikonalNLO} we have arranged the numerators to satisfy color-kinematic duality, in the sense that the eikonal can be decomposed into form~\eqref{eq:dcEikonalYM}.
At NLO, the BCJ double copy kernel and the corresponding array of color factors are the same as the spinless case given in \eqref{eq:BCJkernelNLO} and \eqref{eq:colorNLOnew}, respectively,
whereas the array of numerators read
\begin{align}
    \label{eq:numeratorsNLO}
    \mathcal{N}_j = \left( n^{(123)} \qquad n^{(231)} \qquad n^{(312)} \qquad n^{(0)} \right).
\end{align}
As stated in subsection~\ref{subsec:wqftdoublecopy}, such a decomposition is always possible since the double copy kernel~\eqref{eq:BCJkernelNLO} is invertible.
Moreover, we note that the kinematic numerators are uniquely fix by the ``partial eikonals'' $\mathcal{A}^{(i)}$,
\begin{align}
    \label{eq:NandA}
    \mathcal{A}^{(i)} = \mathcal{K}_{ij} \mathcal{N}_j
    \qquad \Rightarrow \qquad
    \mathcal{N}_j = (\mathcal{K}^{-1})_{ji} \mathcal{A}^{(i)}.
\end{align}
This is in agreement with its quantum counterpart - the 6-point amplitude for 3 distinguishable quark-antiquark pairs in QCD~\cite{Johansson:2015oia}.

\subsection{Leading Yang-Mills radiation}
\label{subsec:YMradiation}
\begin{figure}[t]
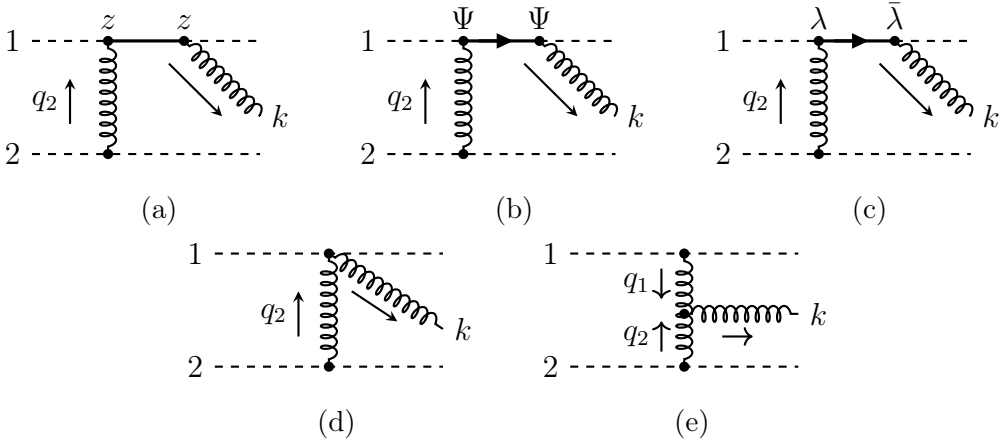

    \centering 
    \begin{subfigure}[b]{0.3\textwidth}
        \centering 
        \loradz{z}{black}{z}
        \caption{}
        \label{kf}
    \end{subfigure}
    \begin{subfigure}[b]{0.3\textwidth}
        \centering
        \loradone{\Psi}{black}{\Psi}
        \caption{}
        \label{cf}
    \end{subfigure}
    \begin{subfigure}[b]{0.3\textwidth}
        \centering
        \loradone{\lambda}{black}{\bar{\lambda}}
        \caption{}
        \label{sf}
    \end{subfigure}
    \\
    \begin{subfigure}[b]{0.3\textwidth}
        \centering 
        \loradYMTwo
        \caption{}
        \label{sb}
    \end{subfigure}
    \begin{subfigure}[b]{0.3\textwidth}\label{sc}
        \centering 
        \loradYMThree
        \caption{}
    \end{subfigure}
    \caption{Topologies contributing to the leading Yang-Mills radiation for spinning worldlines.
        For (\ref{sf}), we also need to include the crossed diagrams that can be obtained by simply reversing the arrows on the worldline.
        For (\ref{kf})-(\ref{sb}), we also need to include diagrams with $1$ and $2$ exchanged.
    }
    \label{fig1}
\end{figure}

Let us now use the Feynman rules to evaluate the leading order ($O(g^{3})$) Yang-Mills radiation emitted in the process of the scattering of two colored and spinning particles.
Similarly to what has been done for the gravitational radiation in \eqref{grad}, we define the Yang-Mills radiation as 
\be 
{\cal R}^{a}_{\mu}\left(\sigma\right)=\la i k^{2}A^{a}_{\mu}(k)\ra_{\textrm{WQFT}} \big|_{k^{2}=0}=  {\cal Z}_{\textrm{YM}}^{-1}\int {\cal D} A  \, e^{i S_{\textrm{YM}}}\, \int \left(\prod_{j=1}^2 \, {\cal D}X_j \, e^{i S[X_j;A]} \right) ik^{2}A^{a}_{\mu}(k)
\ee
where the external gluon has to be considered on-shell while $\sigma=(k,b_{1,2},p_{1,2},\s_{1,2})$ packages all of the scattering data, including the external momentum of the radiated gluon.
The diagrams contributing at the leading order to the radiation are shown in Fig.\ref{fig1}.
Using the Feynman rules defined in the last subsection~\ref{subsec:YMFeyn}, we can get the radiation straightforwardly
\ba
\label{eq:LOradiatonYM}
{\cal R}^{a\mu }\left(\sigma\right) = i g^{3} \int  \dd \mu_{1, 2}(k)
\Bigg(
&\tilde{C}^{a}_{1}\left( -\frac{k \cdot q_{2}}{q_{2}^{2}\omega_{1}^{2}} n^{\mu}_{0} + \frac{1}{q_{2}^{2}\omega_{1}} n^{\mu}_{1}	\right) 
+ \tilde{C}^{a}_{2}\left( -\frac{k \cdot q_{1}}{q_{1}^{2}\omega_{2}^{2}} \bar{n}^{\mu}_{0} +\frac{1}{q_{1}^{2} \omega_{2}} n^{\mu}_{1}	\right) \\
+&\tilde{C}^{a}_{3}\left( \frac{1}{q_{2}^{2} \omega_{1}} n^{\mu}_{0} +\frac{1}{ q_{1}^{2} \omega_{2}} \bar{n}^{\mu}_{0} + \frac{2}{q_{1}^{2}q_{2}^{2}} n_{1}^{\mu}			\right)
\Bigg),
\ea 
where in this case, we have ${\omega}_1 = p_1 \cdot q_2$, and ${\omega}_2 = -p_2 \cdot k$, which are consistent with \eqref{eq:omega_i} upon identifying $k \to -q_3$.
For convenience, we also define the color factors as
\begin{align}
    \tilde{C}_{1}^{a}=C_{1}\cdot C_{2}\, C_{1}^{a}, \quad
    \tilde{C}_{2}^{a}=C_{1}\cdot C_{2}\, C_{2}^{a}, \quad
    \tilde{C}_{3}^{a}=f^{abc}C_{1}^{b}C_{2}^{c},
\end{align}
and the numerators as
\ba
n_{0}^{\mu} =& \gamma\, p_1^\mu
    + i \gamma\, (k \cdot \s_1)^\mu
    -i p_1^\mu (q_2 \cdot \s_1 \cdot p_2 - q_2 \cdot \s_2 \cdot p_1)
    -i p_1 \cdot q_2 (p_2 \cdot \s_1)^\mu \\
    &- q_2 \cdot \s_2 \cdot p_1 (k \cdot \s_1)^\mu
    - q_2 \cdot \s_1 \cdot \s_2 \cdot q_2\, p_1^\mu 
    + p_1 \cdot q_2 (q_2 \cdot \s_2 \cdot \s_1)^\mu \\
n_{1}^{\mu} =& 
    \gamma q_{2}^{\mu} 
    - p_1 \cdot q_2\, p_2^{\mu} + p_2 \cdot q_1\, p_1^\mu
    + q_{1}\cdot \s_{1}\cdot \s_{2} \cdot q_{2}\, q_{2}^{\mu}
    + \big[ i p_1\cdot \s_2 \cdot q_2\, q_1^\mu \\
      &-i q_{1} \cdot \s_{2} \cdot q_{2}\, p_1^{\mu}
      +i p_1 \cdot q_2 (q_{2}\cdot \s_{2})^{\mu}
      + q_{1} \cdot \s_{2} \cdot q_{2} (q_{1} \cdot \s_{1})^{\mu}
      - (1 \leftrightarrow 2) \big]
\ea
with $\bar{n}_{0} = n_{0}|_{1 \leftrightarrow 2}$.
We can see that the leading-order radiation can be obtained from the NLO eikonal \eqref{eq:YMeikonalNLO} by cutting off worldline 3 and putting the external gluon on shell.
Specifically, we send $q_3 \to -k, \s_3 \to 0$, strip off $C_3^a$ and $p_3^\mu$, multiply the result by $ -ik^2/g$, and finally set the outgoing momentum $k^\mu$ on-shell.
We stress that in \eqref{eq:LOradiatonYM}, we have already arranged the numerators to satisfy color-kinematics duality, in the sense that the Yang-Mills radiation can be written in a similar form as \eqref{eq:dcEikonalYM},
\be \label{CKD}
{\cal R}^{a\mu}\left(\sigma\right) = ig^{3} \int \dd\mu_{1,2}(k)\, \sum_{i,j}\tilde{C}^{a}_{i}{\cal K}_{ij} \mathcal{N}^{\mu}_{j},
\ee
where the array of numerators is
\begin{align}
    \mathcal{N}^{\mu} = ( n^{\mu}_{0} \quad \bar{n}^{\mu}_{0} \quad n^{\mu}_{1} )
\end{align}
and the double copy kernel for the LO radiation reads
\be 
{\cal K}_{ij} = \begin{pmatrix}
    -\frac{k\cdot q_{2}}{q_{2}^{2} \omega_{1}^{2}} & 0 & \frac{1}{q_{2}^{2}\omega_{1}}\\
    0 & -\frac{k\cdot q_1}{q_{1}^{2} \omega_{2}^{2}} & \frac{1}{q_{1}^{2}\omega_{2}}  \\
    \frac{1}{q_{2}^{2}\omega_{1}} & \frac{1}{q_{1}^{2}\omega_{2}} & \frac{2}{q_{1}^{2} q_{2}^{2}}
\end{pmatrix},
\ee
which is obtained by considering the same diagrams as in Fig \ref{fig1} but in a theory of scalar worldlines interacting through the bi-adjoint scalar field \cite{Shi:2021qsb, Bastianelli:2021rbt}.
Noticeably, gauge invariance implies the Ward idnetity ${\cal K}_{ij}(k_{\mu} \mathcal{N}^{\mu}_{j}) =0$, such that the double copy kernel has zero determinant, i.e., it is not an invertible matrix in the usual algebraic sense.%
\footnote{
    Instead the kernel can be inverted in the sense of pseudo-inverse matrices \cite{Boels:2012sy}, so to allow for a KLT-like double copy.
}


\section{Dilaton-Gravity on the worldline}\label{secDG}
Here we work out the coupling of the ${\cal N}=2$ SUSY particle to the dilaton and graviton. The free model
was first proposed by \cite{Brink:1976sz}, then, it was consistently coupled to gravity by Bastianelli, Benincasa, and Giombi in
\cite{Bastianelli:2005vk,Bastianelli:2005uy}.
We start by describing the quantization of the worldline particle and, then,
by suitably deforming the SUSY charges of the free theory, we are able to couple the dilaton to such a worldline particle, preserving supersymmetry.
Using the generators of the first class algebra as constraints on the wavefunctions propagated by the worldline, we show that our deformation, in $D=4$, reproduces the results in \cite{Bautista:2019evw}, where the dilaton is sourced to a vector field only in the mass term.
Then, we move to the construction of the WQFT action, which allows us to write Feynman rules for the theory.

\subsection{SUSY in the sky with dilaton}
We start by describing the free theory, building on the results from the ${\cal N}=1$ model.
Let us consider two independent free SUSY charges from two copies of the ${\cal N}=1$ model.
We label the SUSY charges as
\be
Q_{L} = \psi_{L}^{\mu}P_{\mu} - m \theta_{L},\hskip.4cm  
Q_{R} = \psi_{R}^{\mu}P_{\mu} - m \theta_{R}
\ee
 with their related real fermionic variables carrying the same label.
 What we see is that, upon defining complex Grassmann variables
  \be \label{complexG} 
 \psi^{\mu}= \frac{1}{\sqrt{2}}(\psi^{\mu}_{L} - i\psi^{\mu}_{R}), \hskip.5cm 
 \theta^{\mu}= \frac{1}{\sqrt{2}}(\theta^{\mu}_{L} - i\theta^{\mu}_{R})
 \ee
we are able to identify a new SUSY charge $Q = \frac{1}{\sqrt{2}}(Q_L-i Q_R)= \psi^{\mu} P_{\mu}-m\theta $ alongside with its complex conjugate. We can gauge the charges so to write the free ${\cal N}=2$ phase space action as
\be
 S_{\mathrm{ph}} = - \int_{0}^{1} \dd\tau \left( \dot{x}^{\mu} P_{\mu} 
   + i \bar{\psi}_{\mu} \dot{\psi}^{\mu} 
   -i \bar{\theta} \dot{\theta}	
   -\frac{e}{2} P^{2} 
   -i \bar{\chi}Q 
   -i \chi \bar{Q} + a (J-s)\right),
\ee
where $\chi, \bar{\chi}$ are Grassmann-valued Lagrange multipliers gauging the supersymmetry. 
We also gauge a $U(1)$ symmetry of the complex Grassmann variable through the gauge field $a(\tau)$ and the current $J= \bar{\psi}_{\mu}\psi^{\mu}-\bar \theta \theta$, with $s$ being the Chern-Simons integer parameter.
This way we have a R-symmetry on the worldline, under which the Grassmann variables and the $U(1)$ gauge field transform as follows
\be \label{R-symmetry}
\delta \psi^\mu =- i\alpha \psi^\mu, \hskip.3cm \delta \psi^\nu = i \alpha \bar{\psi}^\nu, \hskip.3cm \delta a = \dot{\alpha}
\ee
with $\alpha$ being a gauge parameter.
 The Poisson brackets between worldline variables are defined as
 \be
 \lbrace x^{\mu}, P_{\nu}\rbrace= \delta^{\mu}{}_{\nu}, \hskip.3cm \lbrace \psi^{\mu},\bar \psi_{\nu}  \rbrace = -i \delta^{\mu}{}_{\nu}, \hskip.3cm \lbrace \theta,\bar{\theta}\rbrace = i.
\ee
Let us for a moment focus on the quantization of such a free particle.
Firstly one has that the algebra is of first class, namely
\be
 \lbrace Q,\bar Q \rbrace = -2i H,\hskip.3cm \lbrace Q,J \rbrace =i Q, \hskip.3cm  \lbrace \bar{Q},J \rbrace =-i \bar Q
\ee
with all of the remaining brackets vanishing. 
This allows to use the equations of motion for the gauge fields as constraints at the quantum level.
From the above Poisson brackets, we can implement $\psi\sim \hat{\psi},\bar{\psi}\sim \partial/\partial \psi$ ,\,
$\theta\sim \hat{\theta},\bar \theta \sim -\partial/\partial\theta$ and use a coherent state basis to expand a generic wavefunction of the Hilbert space as
\ba \label{wf}
\Phi(x,\psi,\theta) &= F(x) + F_{\mu}\psi^{\mu} + \frac{1}{2}F_{\mu \nu}(x)\psi^{\mu}\psi^{\nu}+\hdots \frac{1}{D!}F_{\mu_{1} \cdots \mu_{D}}(x)\psi^{\mu_{1}}\hdots \psi^{\mu_{D}}\\
&+im\theta \left( W(x) + W_{\mu}(x)\psi^{\mu}+\hdots \frac{1}{D!}W_{\mu_{1}\cdots \mu_{D}}(x)
\psi^{\mu_{1}}\hdots \psi^{\mu_{D}}\right)\;.
\ea
Thus, we clearly see that the worldline model is propagating totally antisymmetric tensor fields that, once worked out the first class algebra at the quantum level will be identified as $p-$forms.
However, in this way, the model is propagating all of the $p-$forms.
To project on a specific sub-space we need to use the equation of motion of $a(\tau)$ as a quantum constraint on the wave function
\be
\frac{\delta S_{\mathrm{ph}}}{\delta a} = 0 \quad\to\quad 
\left( \hat{J} -s \right) \Phi(x,\psi,\theta) = \left(\psi^{\mu}\frac{\partial}{\partial \psi^{\mu}} + \theta \frac{\partial}{\partial \theta} 	-s	\right)\Phi(x,\psi,\theta) = 0
\ee
such that picking an arbitrary value of $s$, allows to project on a $s$-form and the related $(s\!-\!1)-$form gauge field, namely
\be
\Phi_{s}(x,\phi,\theta) = \frac{1}{s!} F_{\mu_{1}\cdots \mu_{s}}(x)\psi^{\mu_{1}} \cdots \psi^{\mu_{s}} + \frac{im\theta}{(s-1)!}W_{\mu_{1}\cdots \mu_{s-1}}(x)\psi^{\mu_{1}}\cdots \psi^{\mu_{s-1}}\;.
\ee
Then, acting with the Hamiltonian $H = i/2 \lbrace Q,\bar{Q}\rbrace = p^{2}/2$ on the selected wave function
gives the mass-shell condition on each component of the wave function, while the SUSY charges $\hat{Q},\hat{\bar{Q}} |\Phi_{s}\ra =0$ constraints give Bianchi identity on the field strength, transversality condition on the $(s\!-\!1)-$form, alongside with the Proca equation of motions for the $s-$form.
Particularly, in $D=4$ one can see that the model is propagating a massive vector boson and a scalar field (after dualization of the massive forms), exactly as observed in \cite{Johansson:2019dnu} from a field theory viewpoint, when studying the double copy of QCD.
On the classical side, this is in agreement with that the classical double copy of QCD amplitudes leads to quadratic effects in the black hole spin \cite{Maybee:2019jus,Guevara:2019fsj} which, on the worldline, can only be accounted by a model having at least two pairs of real Grassmann variables.

Let us now move to the coupling with the dilaton-gravity background.
From a field theory viewpoint we expect such a worldline model to describe $p-$forms in a curved space coupled to the dilaton.
Inspired by the results in \cite{Bonezzi:2020jjq}, where the ${\cal N}=4$ worldline SUSY model has been used to propagate the supergravity spectrum after quantization, we deform the SUSY charges as
\ba \label{SUSYQ}
Q &= e^{-\kappa \phi} e_{a}^{\mu}\psi^{a}\left( P_{\mu}-i \Sigma_{\mu cd}\bar{\psi}^{c}\psi^{d}		\right) -m\theta \\
\bar{Q} &= e^{-\kappa \phi} e_{a}^{\mu}\bar{\psi^{a}}\left( P_{\mu}-i \Sigma_{\mu cd}	\bar{\psi}^{c}\psi^{d}	\right) -m\bar{\theta}.
\ea
Since now we are in a curve space, it is necessary to differentiate the local flat tangent space, denoted by the Latin indices $a, b, c, \dots$, from the usual covariant curve space denoted by the Greek indices $\mu, \nu, \rho, \dots$.
They are related via the vielbein $e^{a}_{\mu}$ defined as $\eta_{ab} e^a_\mu e^b_\nu = g_{\mu\nu}$.
For example, we have $\psi^\mu = e^\mu_a \psi^a$.
Due to the presence of the dilaton, we use $\Sigma_{\mu a b}=\omega_{\mu ab}-2\kappa \partial_{[a}\phi \, e_{b]\mu}$ as a modification of the spin connection $\omega_{\mu}{}^{a b} = e_\nu^a (\partial_\mu e^{\nu b} + \Gamma^{\nu}{}_{\mu\rho} e^{\rho b})$.
Let us now go through this deformation.
The Poisson bracket $\lbrace Q,Q\rbrace $ identically vanishes when requiring our manifold to be torsionless and upon invoking Bianchi identity on the Riemann tensor. Indeed, the deformation is designed such that this Poisson bracket would deliver the torsion and the Riemann tensor in what is known as the Einstein frame.
The same happens for the bracket $\lbrace \bar{Q},\bar{Q}\rbrace $. More details on this can be seen in Appendix \ref{appA}.
This means that our coupling preserves supersymmetry, thus allowing for a consistent quantization of the model.

Let us inspect the quantum theory implemented by the worldline particle using the SUSY charges as constraints on the wave function \eqref{wf}.
Choosing the Chern-Simons parameter $s=2$ so to propagate a massive vector boson and its related field strength
\be 
\Phi_{2}(x,\psi,\theta) = \frac{1}{2}F_{\mu\nu}(x)\psi^\mu \psi^\nu +im \theta W_\mu (x) \psi^\mu
\ee
and acting on such a state with the SUSY charges \eqref{SUSYQ} one gets
\be \label{eqn}
\hat{Q}^\dagger|\Phi_{2}\ra =0 \to
\begin{cases}
\nabla^\mu F_{\mu\nu} =m^2 e^{\kappa \phi}W_\nu\\
 \nabla_{\mu }W^{\mu}=0
 \end{cases}
 \hskip.4cm \hat{Q} |\Phi_{2}\ra = 0 \to
 \begin{cases}
\nabla_{[\mu}F_{\nu\rho]} = 0 \\
e^{\kappa \phi}F_{\mu\nu} = 2 \nabla_{[\mu}W_{\nu]}
 \end{cases}
\ee
in agreement with the equations of motion from the $D=4$ limit of the theory $\frac{1}{2}\otimes \frac{1}{2}$ in \cite{Bautista:2019evw}, where the dilaton is only sourced to the mass term for the vector boson.
Working out the SUSY algebra from the SUSY charges in \eqref{SUSYQ}, whose details can be found in Appendix \ref{appA}, one is able to write down the following WQFT action for a massive point particle coupled to dilaton-gravity
\be \label{DA}
S_{\mathrm{pm}} = \int_{-\infty}^{\infty} \dd \tau 
\left( -\frac{1}{2}e^{2\kappa \phi}g_{\mu\nu}\dot{x}^\mu \dot{x}^\nu  
-i \bar{\psi}_a \dot{\psi}^a 
+\frac{1}{2} \Sigma_{\mu}{}^{ab} \dot{x}^\mu S_{ab}
+\frac{1}{8} e^{-2\kappa \phi}{\cal R}_{abcd}S^{ab}S^{cd}\right)
\ee
where in this $\mathcal{N}=2$ case the spin tensor is defined as $S^{ab}= -2i \bar{\psi}^{[a}\psi^{b]}$, and the deformed Riemann tensor in the flat space is 
\begin{align}
    \label{eq:deformedRabcd}
    \mathcal{R}_{abcd} =& e_a^\mu e_b^\nu
    \left( e_c^\rho e_d^\sigma R_{\mu\nu\rho\sigma}
      -4\kappa \nabla_{[\mu } \nabla_{[c} \phi e_{d]\nu]} 
      +2\kappa^2 \left(2 \partial_{[c} \phi \partial_{[\mu} \phi e_{\nu] d]}	    
        -\partial^2 \phi e_{[c[\mu}e_{\nu]d]} \right)
    \right).
\end{align}

Let us conclude this section by giving a comment on the above results.
First, one can notice that the action \eqref{DA} is really a Weyl rescaling $g_{\mu\nu} \to e^{2\kappa \phi} g_{\mu\nu}$ of the action of a $\mathcal{N}=2$ particle coupled to pure gravity used in \cite{Jakobsen:2021zvh}.
This is a direct consequence of that even the deformation of the SUSY charges \eqref{SUSYQ} is a Weyl rescaling of the one used in \cite{Bastianelli:2005uy, Jakobsen:2021zvh} to couple the model to gravity. 
This nice property implies that the dilaton can be completely disentangled from the worldline by reversing the Weyl transformation, yielding an action in the string frame.
As we will see in the next subsection, the Feynman rules will be much simpler and less Feynman diagrams will be needed to compute the eikonal and the radiation.

\subsection{WQFT expansion and Feynman rules}
\label{subsec:FeynRulesGR}
As explained before, in order to simplify as much as possible the perturbative expansion, we move back to the string frame by performing the Weyl rescaling
\be
 g_{\mu\nu} = e^{-2\kappa \phi} \tilde{g}_{\mu\nu},
\ee
which allows to rewrite the worldline action \eqref{DA} as
\begin{align}
    \label{eq:WLstringframe}
    S_{\mathrm{pm}} = \int_{-\infty}^{\infty} \dd\tau \bigg(
    -&\frac{1}{2} 
    \tilde{g}_{\mu\nu}\dot{x}^\mu \dot{x}^\nu - i \bar{\psi}_a \dot{\psi}^a 
    + \frac{1}{2} \dot{x}^\mu \tilde{\omega}_{\mu}{}^{a b} S_{ab}
    +\frac{1}{8} \tilde{R}_{abcd}S^{ab}S^{cd}
    \bigg),
\end{align}
where objects in the string frame are tilded.
The spin connection and the Riemann tensor are given in terms of the tilded quantities,
\begin{align}
\tilde{\omega}_{\mu}{}^{a b} &= \tilde{e}_\nu^a (\partial_\mu \tilde{e}^{\nu b} + \Gamma^{\nu}{}_{\mu\rho} \tilde{e}^{\rho b})\\
\tilde{R}_{abcd} &= 2 \tilde{e}^{\mu}_{c} \tilde{e}^{\nu}_{d} \left(\partial_{[\mu} \tilde{\omega}_{\nu] a b}+\tilde{\omega}_{[\mu a}{}^f \tilde{\omega}_{\nu] f b}\right).
\end{align}

In the bulk, we expect to probe the dilaton-gravity sector of the $\mathcal{N}=0$ supergravity, known to arise from the double copy of pure Yang-Mills, and the action reads
\begin{align}
    \label{eq:SdgString}
    S_{\mathrm{dg}} = -\frac{2}{\kappa^2} \int d^D x 
     \sqrt{-\tilde{g}(x)} e^{-2 \phi} \left(
    \tilde{R} + 4 \tilde{g}^{\mu\nu} \partial_\mu \phi \partial_\nu \phi 
    \right).
\end{align}
We find it convenient to expand the metric as
\begin{align} \label{gexp}
    \tilde{g}_{\mu\nu} = e^{\kappa h_{\mu\nu}} = \eta_{\mu\nu} + \kappa h_{\mu\nu} + \frac{\kappa^2}{2} h_{\mu\rho}h^{\rho}{}_\nu + \dots,
\end{align}
where for perturbative quantities such as $h_{\mu\nu}$ the Greek indices are now raised and lowered by the flat metric $\eta_{\mu\nu}$.
In this expansion, to decouple the kinetic terms of $\phi$ and $h_{\mu\nu}$, we perform a field redefinition $\phi \to \tilde{\phi} + \frac{1}{4} h_\mu{}^\mu$, and add to the bulk action the following gauge fixing term
\begin{align}\label{ngf}
    S_{\mathrm{gf}} = \int \dd^D x \left( \partial_{\nu}h^{\mu \nu} + 2 \partial^{\mu }\tilde{\phi} \right) 
    \left( \partial_{\nu}h_{\mu}{}^{\nu} + 2 \partial_{\mu }\tilde{\phi} \right).
\end{align}
In the end, the gauge-fixed dilaton-gravity action simply reads 
\begin{align}
    \label{eq:SdgStringExpand}
    S_{\mathrm{dg}} = \int \dd^D x
      \bigg(& \frac{1}{2} \partial_{\rho}h_{\mu\nu} \partial^{\rho}h^{\mu\nu}
      -4 \partial_{\mu}\tilde{\phi} \partial^{\mu}\tilde{\phi} \nonumber \\
      &- \kappa \Big( \frac{1}{2} h^{\mu \nu} \partial_{\mu}h^{\rho\sigma} \partial_{\nu}h_{\rho\sigma}
       - h^{\mu\nu} \partial_{\nu}h_{\rho\sigma} \partial^{\sigma}h_{\mu}{}^{\rho} \Big) \bigg)
      + \mathcal{O}(\kappa^2),
\end{align}
where we have neglected interaction terms involving the dilaton $\tilde{\phi}$.
Thus, we can write down the WQFT partition function for $n$ worldlines as follows
\be
{\cal Z}_{\mathrm{dg}} = \int D[h_{\mu\nu},\tilde\phi] \,e^{i S_{\mathrm{dg}}[h,\tilde\phi]} \, 
\int \prod_{k=1}^n {\cal D}X_{k}\, e^{i S_{\mathrm{pm}}[X_{k};h]}
\ee
with the worldline variables $X_{k} = (x_{k},\psi_{k},\bar{\psi}_{k})$ and the path integral measure defined as ${\cal D}X_{k} = Dx_{k}D\psi_{k}D\bar{\psi}_{k}$, where we implicitly included the Lee-Yang ghost term as introduced in \eqref{eq:LYmeasure}.
Let us move now to the Feynman rules needed for our calculations.
In this way, the Feynman graviton propagator turns out to be
\be \label{GRprop}
\raisebox{-2mm}{\grprop} = \frac{i}{2 q^{2}}\left(	\eta_{\mu\rho}\eta_{\nu\sigma}+\eta_{\mu\sigma}\eta_{\nu\rho}	\right)\;.
\ee
From the second line in \eqref{eq:SdgStringExpand}, we can extract the three-point graviton vertex, which turns out to be extremely simple and can be written as
\be
\raisebox{-10mm}{\vhhh}=
     \,\frac{i\kappa }{2}  q_1^{\alpha} q_2^{\beta} \eta^{\rho\mu} \eta^{\sigma\nu}
    -i \kappa q_1^{\beta} q_2^{\nu} \eta^{\mu\sigma} \eta^{\rho\alpha} +\textrm{S}_{3}(1,2,3), \nonumber
\ee
where $\textrm{S}_{3}(1,2,3)$ is the set of all of the permutations of the list $(1,2,3)$ labeling the Lorentz indices and momenta of the external gravitons.
We have also implicitly symmetrize in $(\mu, \nu), (\rho, \sigma), (\alpha, \beta)$, separately.
Moving on to the worldline action, we see that, in the string frame, 
the leading PM expansion of the action generates the same Feynman rules given in \cite{Jakobsen:2021zvh}, as a consequence of \eqref{gexp}.
The sub-leading expansion reads as
\begin{align}
    S_{\mathrm{pm}} =& \int d^D x
    \bigg(- \frac{\dot{x}^2}{2} 
    + \kappa \Bigl(-\frac{1}{2} h_{\mu \nu} \dot{x}^{\mu} \dot{x}^{\nu} 
      + \frac{1}{2} S^{\nu \rho} \dot{x}^{\mu} \partial_{\rho}h_{\mu \nu}
      - \frac{1}{4} S^{\mu \nu} S^{\rho \sigma} \partial_{\sigma}\partial_{\nu}h_{\mu\rho} \Bigr) \nonumber \\
    &+ \kappa^2 \Bigl(
      - \frac{1}{4} h_{\mu}{}^{\rho} h_{\nu\rho} \dot{x}^{\mu} \dot{x}^{\nu}
      - \frac{1}{8} h^{\nu\rho} S_{\nu}{}^{\sigma} \dot{x}^{\mu} \partial_{\mu}h_{\rho\sigma}
      + \frac{1}{4} h^{\nu\rho} S_{\nu}{}^{\sigma} \dot{x}^{\mu} \partial_{\rho}h_{\mu\sigma}
      + \frac{1}{4} h_{\mu}{}^{\nu} S^{\rho\sigma} \dot{x}^{\mu} \partial_{\sigma}h_{\nu\rho} \nonumber \\
      &\quad - \frac{1}{8} S^{\mu\nu} S^{\rho\sigma} \partial_{\rho}h_{\mu}{}^{\lambda} \partial_{\sigma}h_{\nu\lambda}
      - \frac{1}{8} S^{\mu\nu} S^{\rho\sigma} \partial_{\nu}h_{\mu}{}^{\lambda} \partial_{\sigma}h_{\rho\lambda}
      - \frac{1}{4} h^{\mu\nu} S_{\mu}{}^{\rho} S^{\sigma\lambda} \partial_{\lambda}\partial_{\nu}h_{\rho\sigma} \nonumber \\
      &\quad + \frac{1}{4} S^{\mu\nu} S^{\rho\sigma} \partial_{\sigma}h_{\nu\lambda} \partial^{\lambda}h_{\mu\rho} 
      - \frac{1}{16} S^{\mu\nu} S^{\rho\sigma} \partial_{\lambda}h_{\nu\sigma} \partial^{\lambda}h_{\mu\rho} \Bigr)
    \bigg) + \mathcal{O}(\kappa^3).
\end{align}
which allows to write down the vertex with the emission of two gravitons from the worldline
\begin{align}
     \raisebox{-10mm}{\vtwoGR} =
     -&\frac{1}{16}i \kappa^2 \Big(\!
     4 p^{\mu } p^{\rho } \eta^{\nu \sigma }
     - 4i q_{2}^{\nu } p^{\rho } S^{\mu \sigma }
     + 4i p^{\mu } (q_2 \cdot S)^{\sigma} \eta^{\nu \rho }  \\[-1.3em]
     &+ 2i (p \cdot q_2) S^{\mu \sigma } \eta^{\nu \rho }  
     + 4 q_{2}^{\nu } S^{\mu \rho } ((q_1\!+\! q_2) \ccdot S)^{\sigma }
     - (q_1 \ccdot q_2) S^{\mu \rho } S^{\nu \sigma } \nonumber \\
     &- 2 S^{\mu \rho } (q_1 \cdot S \cdot q_2) \eta^{\nu \sigma } 
     - 2 (q_1 \cdot S)^{\mu} (q_2 \cdot S)^{\rho } \eta^{\nu \sigma } \nonumber
     \Big) + (1 \leftrightarrow 2).
\end{align}
Again, we need to symmetrize the indices $(\mu,\nu)$ and $(\rho,\sigma)$, separately.
Thus, what we see is that our field redefinition combined with the gauge fixing term \eqref{ngf} allows to disentangle the dilaton from the graviton kinetic term, to drastically simplify the three graviton vertex, and to decouple the dilaton from the worldline action, such that, for our purposes, we should only focus on the interaction concerning the graviton.

\section{Double copy of spinning worldlines}
\label{secDC}
In this section we show how to double copy classical spinning particles, preserving SUSY and R-symmetry on the worldline.
The double copy procedure for multiplying the numerators is the same as the spinless case~\eqref{eq:dcEikonal}.
However, the spin degree of freedom needs to be dealt with carefully.
The spins carried by the two copies, referred to as ``left'' and ``right'', will be labeled as $L$ and $R$, respectively.
The sum of these two spins will yield the spin in the double copy theory.
Moreover, in order to preserve the SUSY and R-symmetry, we will symmetrize the labels $L$ and $R$ for each individual particles in the double copy procedure.
We found that the double copy of the $\mathcal{N}=1$ worldline in YM background leads to the ${\cal N}=2$ particle coupled to DG.
In addition, we also investigate the case where symmetrization of $L$ and $R$ labels is not accounted, in which the B-field will present in the double copy spectrum, breaking SUSY and R-symmetry on the worldline.

\subsection{Double copy of the eikonal phase}
\label{subsec:dcEikonal}
Let us start by considering the simplest case of the leading order eikonal for the scattering of two spinning worldlines in a gauge background.
Firstly, let us clarify how to identify the spin tensor after the double copy.
Given that we expect to probe quadrupole, the underlying WQFT should be constructed with complex Grassmann variables, related to the real ones from the ${\cal N}=1$ model, by \eqref{complexG}. 
This is already selecting the ${\cal N}=2$ model as a free theory. In particular, using \eqref{complexG} 
and \eqref{be-real-psi} this also fixes the background expansion of the complex Grassmann variables of the ${\cal N}=2$ model as
\ba\label{be-dc}
\psi^{\mu}(\tau) &= \frac{1}{\sqrt{2}}\left(\psi^{\mu}_{L}(\tau)-i\psi^{\mu}_{R}(\tau)\right)\\
&= \frac{1}{\sqrt{2}} \left(\zeta^{\mu}_{L}-i\zeta^{\mu}_{R}\right) +\frac{1}{\sqrt{2}} \left(\Psi^{\mu}_{L}(\tau)-i\Psi^{\mu}_{R}(\tau)\right) = \zeta^{\mu} + \Psi^{\mu}(\tau),
\ea
further implying that the Grassmann variables from the two copies should be treated differently, so that we have two real Grassmann variables at the gravity side. 
For convenience, we label them as $L$ and $R$.
We thus propose the following relation between the spin tensor in the gravity and the YM side,
\be \label{spin}
\s^{ab}= \s^{ab}_{L}+\s^{ab}_{R},
\ee
with $\s^{a b}_{L,R} = -i \zeta^{a}_{L,R}\zeta^{b}_{L,R}$, 
so that when written in terms of the complex Grassmann variables in \eqref{be-dc}, the spin tensor is $\s^{ab}=-2i \zeta^{[a}\bar{\zeta}^{b]}$.
This is also consistent with the fact that the Lorentz generator for spin-$s$ particles is basically the sum of $s$ vector representations in the spinor-helicity formalism~\cite{Guevara:2019fsj}.

Let us now move to studying the double copy of the leading eikonal phase.
At LO, it has been calculated in \eqref{eq:YMeikonalLO}.
We take two copies of the numerators, and label the spins by $L$ and $R$, interpreted as labeling the left and a right copy of the ${\cal N}=1$ theory. 
For example, if both particles are labeled with $L$, the numerator reads
\begin{align}
    \mathcal{N}_{L_1 L_2} = \gamma
    - i \big( q_2 \cdot \s_{1,L} \cdot p_2
    - q_2 \cdot \s_{2,L} \cdot p_1 \big)
    - q_2 \cdot \s_{1,L} \cdot \s_{2,L} \cdot q_2\;.
\end{align}
In such a case, following \eqref{eq:dcEikonal},  the naive double copy procedure yields
\be 
\label{naive}
\chi_1^{\mathrm{DC}} \sim \frac{\mathcal{N}_{L_1 L_2} \, \mathcal{N}_{R_1 R_2}}{q_2^2}.
\ee
The question now is how to select a worldline theory generating such numerator and whether it enjoys supersymmetry or R-symmetry, the former being crucial to fix all of the interactions on the wordline, while the latter, allowing to propagate a specific particle among the worldline spectrum.
Let us now discuss the consequences of R-symmetry at the level of classical integrands from the WQFT.
After gauge fixing the $U(1)$ gauge field $a(\tau)$ on the worldline, the theory enjoys the global version of the symmetry \eqref{R-symmetry} on the background parameters $\zeta,\bar{\zeta}$, 
which can be used to check global $U(1)$ invariance of the double copy integrand. 
This amounts to check that the integrand can be expressed entirely in terms of global R-invariant objects i.e. in terms of the spin tensor $\s^{ab}$.

Then, this is signaling that \eqref{naive}, cannot be generated by a R-invariant theory!
To keep $U(1)$ invariance, we perform the replacement above, then symmetrizing over the left and right indices, such that, the double copied eikonal turns out to be
\ba
    \chi_1^{\mathrm{DC}} &= -\frac{\kappa^{2}}{4} \int \dd \mu_{1,2}(0) 
    \frac{\mathcal{N}_{L} \otimes \mathcal{N}_{R}}{q_2^{2}}
\ea
where $\otimes$ defines symmetrization over $L_{1}, R_{1}$ and $L_{2},R_{2}$ separately. This yields to the double copied numerator
\ba\label{dc-sym-num}
\mathcal{N}_{L}\otimes \mathcal{N}_{R} =&
\gamma^2-i \gamma\, q_2\cdot \s_1\cdot p_2+i \gamma\, q_2\cdot \s_2\cdot p_1\\
&-\frac{1}{2}  \left(q_2\cdot \s_1\cdot
   p_2\right){}^2-\frac{1}{2} \left(q_2\cdot \s_2\cdot p_1\right){}^2
   -\frac{1}{2} i \, q_2\cdot \s_1\cdot \s_2\cdot q_2  \, q_2\cdot \s_2\cdot
   p_1\\
   &+\frac{1}{2} i \, q_2\cdot \s_1\cdot  \s_2\cdot q_2 \, q_2\cdot \s_1\cdot p_2+\frac{1}{2} \, q_2\cdot \s_1\cdot p_2 \, q_2\cdot \s_2\cdot
   p_1\\
   &-\frac{1}{2} \gamma \, q_2\cdot \s_1\cdot \s_2\cdot q_2+\frac{1}{4} \left(q_2\cdot \s_1\cdot  \s_2\cdot q_2 \right){}^2
\ea
which is manifestly background R-invariant, given that it can be recast in terms of the ${\cal N}=2$ spin tensor.
We stress that a crucial role in obtaining the above numerator from the double copy of the ${\cal N}=1$ model, is played by the real Grassmann variables $\psi^{\mu}_{L,R}$. 
Indeed, some of the terms one gets vanish once using that $\psi_{L,R}^{2}=0$ and anti-symmetry of the string $\psi^{\mu}_{L,R}\psi^{\nu}_{L,R}$.
For instance, a factor quadratic in $\s_1$ will be simplified as
\begin{align}
    & (q_2\cdot \s_{1,L})^\mu\,  (q_2\cdot \s_{1,R})^\nu
    + (q_2\cdot \s_{1,R})^\mu \, (q_2 \cdot \s_{1,L})^\nu \\
    =&(q_2\cdot \s_{1,L})^\mu\,  (q_2\cdot \s_{1,L})^\nu
    + (q_2\cdot \s_{1,L})^\mu \,  (q_2\cdot \s_{1,R})^\nu \nonumber \\
    +&(q_2\cdot \s_{1,R})^\mu \,  (q_2\cdot \s_{1,L})^\nu 
    + (q_2\cdot \s_{1,R})^\mu \, (q_2 \cdot \s_{1,R})^\nu \nonumber \\
    =& (q_2\cdot \s_{1})^\mu\,  (q_2\cdot \s_{1})^\nu, \nonumber 
\end{align}
where we have used the fact that 
$(q_2\cdot \s_{1,L})^\mu\,  (q_2\cdot \s_{1,L})^\nu \sim q_{2\alpha}q_{2\beta}\zeta_{L}^{\alpha}\zeta_{L}^{\beta}\zeta_{L}^{\mu}\zeta_{L}^{\nu} = 0$ due to the anti-symmetry of the Grassmann variables. The same holds for the right copy of such term.
By inspection, it can be seen that the double copied numerator \eqref{dc-sym-num} is in agreement with the one extracted from the leading eikonal evaluated using the Feynman rules in subsection \ref{subsec:FeynRulesGR}, thus automatically checking SUSY invariance of the double copied eikonal phase.
We note that the Grassmann nature of the spin tensor is critical in matching the double copy eikonal to the direct calculation from dilaton gravity.

\begin{figure}[t]
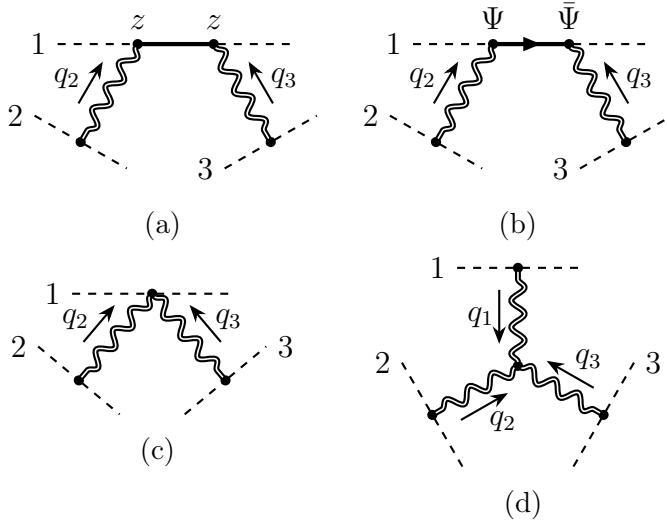

    \centering
    \begin{subfigure}[b]{0.3\textwidth}
        \centering
        \eikonalYMNLOz{graviton}
        \caption{}
        \label{fig:DGeikonalNLOz}
    \end{subfigure}
    \begin{subfigure}[b]{0.3\textwidth}
        \centering
        \eikonalYMNLOpsi{\Psi}{\bar{\Psi}}{graviton}
        \caption{}
        \label{fig:DGeikonalNLOpsi}
    \end{subfigure}
    \\
    \begin{subfigure}[b]{0.3\textwidth}
        \centering
        \eikonalYMNLOcontact{graviton}
        \caption{}
        \label{fig:DGeikonalNLOcontact}
    \end{subfigure}
    \raisebox{-7mm}{
        \begin{subfigure}[b]{0.3\textwidth}
            \centering
            \eikonalYMNLOgluons{graviton}
            \caption{}
            \label{fig:DGeikonalNLO3gluon}
    \end{subfigure}}
    \caption{The diagrams needed for the DG eikonal at NLO with three worldlines.
        For diagrams involving the propagators of $\Psi$ (\ref{fig:DGeikonalNLOpsi}), we also need to include the crossed diagrams that can be obtained by simply reversing the arrows on the worldline.
        We only display diagrams with worldline propagator and contact interaction of particle~1.
    }
    \label{fig:DGeikonalNLO}
\end{figure}

Let us now proceed to the NLO.
At this order, again, the double copy multiplication rule is the same as the spinless case, with the numerators given in \eqref{eq:numeratorsNLO}.
Written with the $\otimes$ symmetrization prescription, the double copy eikonal  \eqref{eq:dcEikonalDG} can be expressed as
\begin{align}
    \chi_2^{\mathrm{DC}} = &-\frac{\kappa^4}{16} \int \dd \mu_{1,2,3}(0) 
        \sum_{i,j} \mathcal{K}_{ij} \mathcal{N}_{i,L} \otimes \mathcal{N}_{j,L} \nonumber \\
        =& -\frac{\kappa^4}{16} \int \dd \mu_{1,2,3}(0)
        \bigg[
        \Big( \frac{2}{q_2^2 q_3^2 \omega_1} n_L^{(123)} \otimes n_R^{(0)}
        + \frac{q_2 \ccdot q_3}{q_2^2 q_3^2 \omega_1^2} n_L^{(123)} \otimes n_R^{(123)}
        + \text{cyclic} \Big) \nonumber \\
        &\hskip3.5cm + \frac{2}{q_1^2 q_2^2 q_3^2} n_L^{(0)} \otimes n_R^{(0)}
        \bigg].
\end{align}
One can check that, with the symmetrization of $L_i$ and $R_i$ labels, this agrees with the direct computation of the Feynman diagrams
(see Fig. \ref{fig:DGeikonalNLO}) from DG using the Feynman rules in subsection~\ref{subsec:FeynRulesGR}.
The final double copy numerators of the NLO eikonal (as well as the radiation) are too lengthy to fit into this article, instead they are provided in an attached ancillary file.

As stated in subsection \ref{subsec:wqftdoublecopy}, the double copy kernel for the NLO eikonal is invertible, so alternatively we can easily perform the double copy in a KLT-like fashion.
Specifically, according to the relation between the ``partial eikonals'' and the kinematic numerators \eqref{eq:NandA}, the eikonal phase at the gravity side of the double copy theory can be re-expressed as
\begin{align}
    \chi_2^{\mathrm{DC}} = -\frac{\kappa^4}{16} \int \dd \mu_{1,2,3}(0) 
    \sum_{i,j} (\mathcal{K}^{-1})_{ij} \mathcal{A}^{(i)}_{L} \otimes \mathcal{A}^{(j)}_{R},
\end{align}
where again, the $\otimes$ denotes the symmetrization of the $L$ and $R$ labels for each individual worldline.
We have checked that this yields exactly the same result as the BCJ double copy.

\subsection{Double copy of the radiation}
\begin{figure}[t]
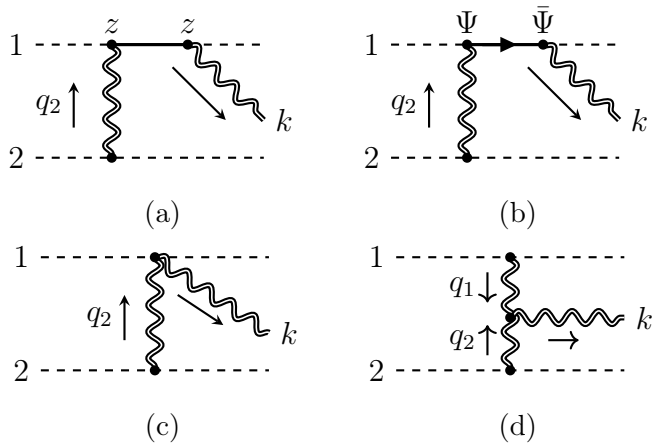

    \centering 
    \begin{subfigure}[b]{0.3\textwidth}
        \centering 
        \loradzGR{z}{black}{z}
        \caption{}
        \label{dgkf}
    \end{subfigure}
    \begin{subfigure}[b]{0.3\textwidth}
        \centering
        \loradoneGR{\Psi}{black}{\bar{\Psi}}
        \caption{}
        \label{dgcf}
    \end{subfigure}
    \\
    \begin{subfigure}[b]{0.3\textwidth}
        \centering 
        \loradGRTwo
        \caption{}
        \label{dgsb}
    \end{subfigure}
    \begin{subfigure}[b]{0.3\textwidth}
        \label{dgsc}
        \centering 
        \loradGRThree
        \caption{}
    \end{subfigure}
    \caption{Topologies contributing to the leading-order gravitational radiation for spinning worldlines.
        For diagrams (\ref{dgcf}), we also need to include the crossed diagrams that can be obtained by simply reversing the arrows on the worldline.
        For (\ref{dgkf})-(\ref{dgsb}), we also need to include diagrams with $1$ and $2$ exchanged.
    }
    \label{fig-rad-GR}
\end{figure}

Let us now move to the double copy of the leading radiation.
We have seen that the radiation can be recasted in a manifest color-kinematic fashion, from the analysis in Sec.\eqref{subsec:YMradiation}. Then, the double copy procedure would require us to replace the color factors $\tilde{C}^{a}_{i }\to \mathcal{N}^{\mu}_{iR}$, the latter being the numerator computed from the right copy of the ${\cal N}=1$ model. However, as pointed out previously, this procedure does not preserve R-symmetry. Then, as for the eikonal phase, we perform the above replacement, then using the $\otimes$ symmetrization prescription on the labels of the two independent copies of the ${\cal N}=1$ particle. This yields the double copied radiation
\ba
{\cal R}^{\mu\nu}(\sigma) &= i\frac{\kappa^{3}}{8}\int \dd \mu_{1,2}(k) \sum_{i,j} {\cal K}_{ij}\,  \mathcal{N}_{i L}^{\mu}\otimes  \mathcal{N}_{j R}^{\mu}\\
&= i\frac{\kappa^{3}}{8}\int \dd \mu_{1,2}(k) \, 
\Bigg(
-\frac{k\cdot q_{2}}{q_{2}^{2}\omega_{1}^{2}} {n}^{\mu}_{0L }\otimes  {n}^{\nu}_{0R } 
+ \frac{2}{q^{2}_{2}\omega_{1}}n^{(\mu}_{1L}\otimes n^{\nu)}_{0R}
- \frac{k\cdot q_{1}}{q_{1}^{2}\omega_{2}^{2}}{\bar{n}}^{\mu}_{0L }\otimes  {\bar{n}}^{\nu}_{0R } \\
 &\hskip3.6cm+ \frac{2}{q_{1}^{2}\omega_{2}}n^{(\mu}_{1L}\otimes{\bar n}^{\nu)}_{0R}
+ \frac{2}{q_{1}^{2}q_{2}^{2}}n^{\mu}_{1L}\otimes n^{\nu}_{1R}
\Bigg) 
\ea
with the shorthand notation $L=L_{1},L_{2}$, $R=R_{1},R_{2}$.
As for the eikonal, so to rewrite the above result in terms of the spin tensor \eqref{spin} one needs to use anti-symmetry of the real Grassmann variables belonging to the left and right copy. One can check that the above radiation correctly reproduces the result obtained by evaluating the diagrams in Fig.\ref{fig-rad-GR} on the gravitational side.

\subsection{B-field in the double copy spectrum}
In the above derivation of the double copy of radiation and eikonal, we mostly relied on the R-symmetry as a guideline. We found that, in this case, the $\otimes$ prescription preserves even SUSY in the double copied integrands. Here we wonder if this is not the case when the R-symmetry is not preserved during the double copy. 
To investigate this, let us study the consequences of the double copy replacement \eqref{naive} which does not preserve R-symmetry in the integrands.
In such a case, given that we cannot use the R-symmetry as a guideline,
we study the double copy features of the three point worldline vertex from the first line in \eqref{LORule}. The replacement \eqref{naive} can be implemented at this level by sending the color factor to a right copy of the vertex, carrying a polarization $\bar{\epsilon}(q)$. Then, we can introduce the NS-NS spectrum by using the expansion
\be
\epsilon_{\mu}(q)\bar{\epsilon}_{\nu}(q) = \epsilon^{h}_{\mu\nu}(q) + \epsilon^{B}_{\mu\nu}(q) + \epsilon\cdot \bar{\epsilon}\left(-\eta_{\mu\nu}+ \frac{2q_{(\mu} r_{\nu)}}{ r\cdot q}	\right)
\ee
with $r^{\mu}$ being a arbitrary null vector.
Separately contracting the double copied vertex with the above expansion and using \eqref{spin} with the SSC, we reproduce the graviton and dilaton vertices which can be checked against \eqref{DA}, with, in addition, the contribution from the $B$ field, whose three point vertex can be reconstructed as
\be \label{Bvertex}
\raisebox{-10mm}{\vaxion} \hskip-.6cm = \frac{i\kappa }{2}e^{i q\cdot b}\deltahat(q\cdot p)\left(
i p^{[\mu} q\cdot(\s_{L} - \s_{R})^{\nu]} + (q\cdot \s_{L})^{[\mu} (q\cdot \s_{R})^{\nu]}
\right)
\ee
which manifestly breaks R-invariance, given the mixing of the left and right copies of the spin tensor.
By studying the structure of such vertex we can reproduce this interaction by deforming the SUSY charges \eqref{SUSYQ} as $Q\to Q - \frac{1}{4}\kappa e^{\kappa \phi}e^{\mu}_{a}\psi^{a}\,H_{\mu cd}T^{cd}$,
with the twisted Lorentz generator $T^{ab} = -i\psi^a_L \psi^b_L +i\psi^a_R \psi^b_R = S^{ab}_L-S^{ab}_R$ as coming from each copies of the ${\cal N}=1$ model.
As it can be checked by working out the SUSY algebra, this deformation breaks supersymmetry, then
showing in our case that, if R-symmetry is not preserved neither SUSY is on the double copied theory.
In addition, given the constraint algebra is not of first class $\lbrace Q,Q\rbrace \neq 0$ we cannot quantize the model as previously, and, particularly, this does not allow to write down a path integral to access classical calculations.
However, using \eqref{DA} as a base theory,  one can write down an effective WQFT action, capturing such effects, namely
\ba \label{EFT}
S = \int_{-\infty}^{\infty} \dd\tau \Bigg(
-&\frac{1}{2}e^{2\kappa \phi}g_{\mu\nu}\dot{x}^\mu \dot{x}^\nu  
-i \bar{\psi}_a \dot{\psi}^a 
+\frac{1}{2} \Sigma_{\mu}{}^{ab} \dot{x}^\mu S_{ab}
+\frac{1}{8} e^{-2\kappa \phi}{\cal R}_{abcd}S^{ab}S^{cd}\\
-&\frac{1}{4} \kappa \dot{x}^\mu H_{\mu a b} T^{ab}
-\frac{1}{8}\kappa \nabla_\mu H_{\nu a b} S^{\mu\nu}T^{ab}
\Bigg)\,.
\ea
This theory agrees up to terms linear in spin with the action proposed by Goldberger and Li in \cite{Goldberger:2019xef} to study the double copy of spinning worldlines by solving the classical equation of motions.
We check that \eqref{EFT} reproduces the leading binary radiation from \cite{Goldberger:2019xef}, confirming their string interpretation of the above action by extending these results up to quadratic order in spin.

\section{Conclusions}
The double copy is a fascinating and powerful tool which allows to study gravitational theories from knowledge of gauge theories. 
At the level of predictions, it allows to boost calculations in a gravitational theory since the perturbative expansion in a gauge theory is much more under control.
Although it was originally formulated in quantum field theory, in the recent years, more and more evidences showed that the double copy can be generalized to the classical level.
In this work, we study WQFT formalism of spinning particles coupled to YM field and to dilaton-gravity, with the spin degrees of freedom characterized by Grassmann valuables.
The former is constructed to feature the $\mathcal{N}=1$ supersymmetry on the worldline, and the latter features the $\mathcal{N}=2$ supersymmetry and the $R$-symmetry.
We generalized the double copy relation proposed in \cite{Shi:2021qsb} to spinning particles, where the relation between the spin tensors at the YM and gravity side is given as \eqref{spin}.
Taking advantages of the anti-symmetry of Grassmann valuables, we recover the eikonal up to $\mathcal{O}(\kappa^4)$ and the radiation at $\mathcal{O}(\kappa^3)$ of the $\mathcal{N}=2$ dilaton-gravity theory from the $\mathcal{N}=1$ YM theory.

Our work sheds light on dealing with classical spin degrees of freedom.
Using the R-symmetry as a guideline on the worldline allows us to write the double copy numerators arising from a WQFT featuring both supersymmetry and R-symmetry. 
Crucially, this makes the constraint algebra of first class, then enabling us to quantize the worldline model, so to get information about the quantum field theory on the double copied side, thus deriving quantum information from classical inputs.
Indeed, starting from the ${\cal N}=1$ SUSY model on the worldline, which propagates Dirac fermions, we identify its classical double copy with the ${\cal N}=2$ model coupled to DG background. 
In addition, given that SUSY is preserved, we quantize the model, finding that in four dimensions, the double copy theory describes a massive vector boson coupled to DG, in such a way that the dilaton is only sourced by the mass term of the vector.
This is in agreement with the results in \cite{Bautista:2019evw, Johansson:2019dnu}.

We stress that a crucial role in the double copy is played by the left and right labels used to distinguish between the two copies of the ${\cal N}=1$ model.
In particular, the symmetrization of the left and right labels allows to keep global $U(1)$ invariance, leading to a theory enjoying R-symmetry at the classical and quantum level, and ruling out the $B$ field from the double copy spectrum.
Further, this is very similar to symmetrizing over the little group indices used when dealing with the double copy of amplitudes written in terms of the massive spinor helicity variables \cite{Arkani-Hamed:2017jhn, Ochirov:2018uyq, Lazopoulos:2021mna}. 
At the level of the double copy spectrum they do essentially the same job.
This similarity is worth being investigated further in future, given that it could lead to a better understanding on how to propagate massive higher spin fields on the worldline, providing a relation between WQFT and spinor helicity formalism, which, in our opinion, seems to be one of the best candidates to deal with higher spin amplitudes.

Another interesting point to be further developed concerns whether the double copy we proposed could help to detect higher spin vertices on the WQFT, given the standard quantum mechanical rules to add spin.
On a different side, with the ${\cal N}=1$ WQFT model in hand, it would worth investigating further the relation between KMOC and WQFT in the case of spinning observables, applying then such result to the calculation of off-shell currents related to hard thermal loops in QCD, extending then the results of \cite{Comberiati:2022ldk}.

\begin{acknowledgments}
We are grateful to F. Bastianelli and R. Bonezzi for enlightening discussions on the quantization of worldline theories.
We thanks G. Mogull and J. Plefka for illuminating discussions concerning the worldline quantum field theory.
The work of FC has been supported by the ``Marco Polo'' fellowship provided by the University of Bologna.
FC would like to thank the Humboldt University for hospitality during the preparation of most of this work.
\end{acknowledgments}

\appendix 
\section{SUSY algebra and DG background}\label{appA}
Here we work out the coupling of the ${\cal N}=2$ SUSY model to the dilaton and gravity, obtained by deforming the free SUSY charges as
\ba \label{SUSYQapp}
Q &= e^{-\kappa \phi} e_{a}^{\mu}\psi^{a}\left(P_{\mu}-i\Sigma_{\mu cd}\bar{\psi}^{c}\psi^{d} \right) -m\theta 
= e^{-\kappa \phi} \psi^a \pi_a -m\theta \\
\bar{Q} &= e^{-\kappa \phi} e_{a}^{\mu}\bar{\psi^{a}} \left(P_{\mu}-i \Sigma_{\mu cd}\bar{\psi}^{c}\psi^{d} \right) -m\bar{\theta}
=e^{-\kappa \phi} \bar{\psi}^a \pi_a -m\bar{\theta} 
\ea
where $\phi$ is the dilaton field and $\Sigma_{\mu c d}=\omega_{\mu cd}-2\kappa\partial_{[c}\phi e_{d]\mu}$.
In the above lines we defined the generalized momentum 
\be 
\pi_a = e_a^\mu \pi_\mu = e_a^\mu \left(P_\mu -i \Sigma_{\mu cd}\bar{\psi}^c \psi^d	 \right),
\ee
where the spin connection $\Sigma_{\mu c d}$ introduced in the deformation, exactly corresponds to the the spin connection written in the Einstein frame, the latter reached through the Weyl rescaling 
\be 
\tilde{e}_{a}^{\mu}= e^{-\kappa \phi}e_a^{\mu} \qquad
\tilde{g}_{\mu\nu} = e^{2\kappa \phi}g_{\mu\nu}
\ee
with tilded objects defined in the string frame.
Next, so to consistently quantize the model and write down a path integral, we need to ensure that the constraint algebra is of first class, particularly that $\lbrace Q,Q\rbrace=\lbrace \bar{Q},\bar{Q}\rbrace=0$. To this aim we first list here the free theory Poisson brackets
\be 
\lbrace x^\mu,P_\nu\rbrace = \delta^\mu{}_\nu \hskip.4cm 
\lbrace \psi^a,\bar{\psi}_b \rbrace=-i \delta^a{}_b\hskip.4cm 
\lbrace \theta ,\bar{\theta}\rbrace = i,
\ee
then, we can evaluate the bracket $\lbrace Q,Q\rbrace$, yielding
\ba  \label{ev}
\lbrace Q,Q\rbrace =& 2
 e^{-\kappa \phi} \psi^a \left(\lbrace \pi_a,e^{-\kappa \phi}\rbrace \psi^b	+e^{-\kappa \phi}\lbrace \pi_a,\psi^b\rbrace 	\right)e_b^{\nu}\pi_\nu +e^{-2\kappa \phi}\psi^a \psi^b \lbrace \pi_a,\pi_b\rbrace  \\
 =&2 e^{-2\kappa \phi} \left(\psi^a\lbrace \pi_a,e^{-\kappa \phi}\rbrace \psi^b e_b^{\nu}	+e^{-\kappa \phi}\psi^a\lbrace \pi_a,\psi^b\rbrace e_b^{\nu}-\psi^a \psi^b e_{[a}^\mu \partial_\mu e_{b]}^\nu	\right)\pi_\nu \\
 &+e^{-2\kappa \phi}e_a^\mu e_b^\nu \psi^{a} \psi^{b} \lbrace \pi_\mu,\pi_\nu\rbrace\\
 =&2 e^{-2\kappa \phi} \left(
 \kappa \partial_a \phi \psi^a \psi^b e_b^\nu -\psi^a \psi^d e_b^\nu\omega_{[a d]}\,^b -\kappa \partial_a \phi \psi^a \psi^b e_b^\nu -\psi^a \psi^b e_{[a}^\mu \partial_\mu e_{b]}^\nu
 \right)\pi_\nu \\
 &+e^{-2\kappa \phi}e_a^\mu e_b^\nu \psi^{a} \psi^{b} \lbrace \pi_\mu,\pi_\nu\rbrace\\
 =&-2 e^{-2\kappa \phi}\left(\omega_{[ab]}\,^d e_d^\nu 
 + e_{[a}^\mu \partial_\mu e_{b]}^\nu  \right) \psi^{a} \psi^{b} \pi_\nu 
 +e^{-2\kappa \phi}e_a^\mu e_b^\nu \psi^{a} \psi^{b}  \lbrace \pi_\mu,\pi_\nu\rbrace \;.
\ea
What we see is that, a direct consequence of our SUSY charge deformation is that, the modification of the spin connection term in \eqref{SUSYQapp} allows us to generate the torsion tensor in the Einstein frame
\be 
T^\nu_{[ab]}=e^{-2\kappa \phi}\left(\omega_{[ab]}\,^d e_d^\nu + e_{[a}^\mu \partial_\mu e_{b]}^\nu  \right)
\ee
which we set to zero assuming our connection is symmetric.
This way, once evaluating the last term in \eqref{ev},  we can recast the bracket as
\begin{gather}
\lbrace Q,Q\rbrace = i \psi^\mu \psi^\nu \bar{\psi}^c \psi^d e^{-2\kappa \phi}{\cal R}_{\mu\nu cd} \\
\text{with} \quad
{\cal R}_{\mu\nu cd} = \left(
R_{\mu\nu cd}
- 4\kappa \nabla_{[\mu }\nabla_{[c} \phi e_{d]\nu]}
+ 2\kappa^2 \big( 2 \partial_{[c}\phi \partial_{[\mu}\phi e_{\nu] d]}	
\!-\! \partial^2 \phi e_{[c[\mu}e_{\nu]d]} \big)
\right)
\end{gather}
where, from the last line, we can read out the Riemann tensor in the Einstein frame. Thus, invoking Bianchi identity and assuming no torsion imply that the above bracket must vanish. A similar calculation holds for $\lbrace \bar{Q},\bar{Q}\rbrace$.
Now one can evaluate the point particle Hamiltonian by using that $\lbrace Q,\bar{Q}\rbrace = -2i H$. The calculation delivers the following Hamiltonian
\be 
H = \frac{1}{2}e^{-2\kappa \phi } \left(g^{\mu \nu}\pi_{\mu}\pi_{\nu}-m^{2}e^{2\kappa \phi }-{\cal R}_{ab cd}\bar{\psi}^{a}\psi^{b}	\bar{\psi}^{c}\psi^{d} \right)
\ee
where the deformed Riemann tensor ${\cal R}_{ab cd}$ is given in \eqref{eq:deformedRabcd}.
Once switching off the dilaton field, $H$ reduces to the Hamiltonian in the pure gravity case~\cite{Jakobsen:2021zvh}.
Now we can write down the worldline action in configuration space.
We consider the phase space action gauging the reparametrization invariance, supersymmetry, and R-symmetry
\be 
S_{\mathrm{ph}}= - \int_{0}^{1}d\tau\left(\dot{x}^{\mu}P_{\mu}+i \bar{\psi}_{a} \dot{\psi}^{a}-i \bar{\theta}\dot{\theta}-eH-i\bar{\chi}Q-i \chi \bar{Q} + a (J-s)\right).
\ee
Then, we eliminate the momentum using the equation of motion
\be
\frac{\delta S_{\mathrm{ph}}}{\delta P_\mu} =0 \to P_\mu = e^{-1} \, e^{2\kappa \phi}\left(
g_{\mu\nu} \dot{x}^\nu -i \chi e^{-\kappa \phi}\bar{\psi}_\mu -i \bar{\chi} e^{-\kappa \phi} \psi_\mu-\frac{e}{2}e^{-2\kappa \phi} \Sigma_{\mu a b}S^{ab}
\right)
\ee
once defining the spin tensor as $S^{ab} = -2i \bar{\psi}^a \psi^b$. Plugging it back in the Hamiltonian and the SUSY charges one is able to write down a configuration space action, ready to be gauge fixed
\ba
S= \int_0^1 d\tau \Big( &-\frac{1}{2}e^{-1} e^{2\kappa \phi}\left(g_{\mu\nu}\dot{x}^\mu \dot{x}^\nu-e^2 m^2 	\right) -i \bar{\psi}_a\dot{\psi}^a +i \bar{\theta}\dot{\theta} +\frac{1}{2} \dot{x}^{\mu}\Sigma_{\mu a b}S^{ab}\\
&+\frac{e}{8}e^{-2\kappa \phi}{\cal R}_{abcd}S^{ab}S^{cd}
+ ie^{-1} e^{\kappa \phi} g_{\mu\nu} \dot{x}^\mu (\bar{\chi} \psi^\nu 
+ \chi \bar{\psi}^\nu)\\
&+im( \chi \bar{\theta}+\bar{\chi}\theta)
-\bar{\chi} \chi e^{-1} \bar{\psi}_a \psi^a
-a (J-s)
\Big)
\ea
which should be gauge fixed according to the topology one would like to evaluate the path integral.
For our classical application, we set $\theta = \bar{\theta}=0$, while using the equations of motion for the gravitinos $(\chi,\bar{\chi})$ and setting them to zero implement the spin supplementary condition. 
In addition, the constraint arising from the gauge field $a(\tau)$ allows to recover the normalization of the spin tensor $S^{\mu\nu}S_{\mu\nu}=2s^{2}$ through the condition $\bar \psi \cdot \psi=s$.

Analogous to the YM case, we fix the einbein by choosing $e = 1/m$, then change the integration boundaries to $(-\infty, \infty)$ by the LSZ reduction procedure.
Upon rescaling the integration variable $\tau\to m\tau$, we obtain the following $\mathcal{N}=2$ worldline action coupled to dilaton-gravity
\be 
S = \int_{-\infty}^{\infty} d\tau \left(-\frac{1}{2}e^{2\kappa \phi}g_{\mu\nu}\dot{x}^\mu \dot{x}^\nu  
-i \bar{\psi}_a \dot{\psi}^a 
-i \dot{x}^\mu \Sigma_\mu{}^{ab} \bar{\psi}_a \psi_b 
+\frac{1}{8} e^{-2\kappa \phi}{\cal R}_{abcd}S^{ab}S^{cd}\right).
\ee

\bibliographystyle{JHEP}

\renewcommand\bibname{References} 
\ifdefined\phantomsection		
\phantomsection  
\else
\fi
\addcontentsline{toc}{section}{References}

\bibliography{dcBib.bib}

\end{document}